\documentclass[a4paper,fleqn]{cas-sc}

\usepackage[authoryear,longnamesfirst]{natbib}
\usepackage{inputenc}
\usepackage{amsmath}
\usepackage{subfig}
\usepackage{amsmath,bm}
\usepackage{makecell}
\usepackage{soul}

\def\tsc#1{\csdef{#1}{\textsc{\lowercase{#1}}\xspace}}
\tsc{WGM}
\tsc{QE}
\tsc{EP}
\tsc{PMS}
\tsc{BEC}
\tsc{DE}

\begin{document}
\let\WriteBookmarks\relax
\def\floatpagepagefraction{1}
\def\textpagefraction{.001}

\shorttitle{H$^2$CGL: Modeling Dynamics of Citation Network for Impact Prediction} 

\shortauthors{G. He et~al.}

\title [mode = title]{H$^2$CGL: Modeling Dynamics of Citation Network for Impact Prediction}
    
\author[1,2]{Guoxiu He}[orcid=0000-0002-1419-7495]
\ead{gxhe@fem.ecnu.edu.cn}
\cormark[1]
\credit{Conceptualization, Methodology, Formal analysis, Investigation, Data curation, Writing - review \& editing}
\affiliation[1]{organization={School of Economics and Management, East China Normal University},
    city={Shanghai},
    postcode={200062}, 
    country={China}}
\affiliation[2]{organization={Institute of AI for Education, East China Normal University},
    city={Shanghai},
    postcode={200062}, 
    country={China}}

\author[1]{Zhikai Xue}
\ead{zkxue@stu.ecnu.edu.cn}
\credit{Conceptualization, Methodology, Software, Validation, Formal analysis, Investigation, Writing - original draft}

\author[3]{Zhuoren Jiang}
\ead{jiangzhuoren@zju.edu.cn}
\affiliation[3]{organization={School of Public Affairs, Zhejiang University},
    city={Hangzhou},
    postcode={310058}, 
    country={China}}
\credit{Conceptualization, Investigation, Writing - review \& editing}

\author[4]{Yangyang Kang}
\ead{yangyang.kangyy@alibaba-inc.com}
\affiliation[4]{organization={Alibaba Group},
    city={Hangzhou},
    postcode={310058}, 
    country={China}}
\credit{Conceptualization, Investigation, Writing - review \& editing}
    
\author[5,6]{Star Zhao}
\ead{xzhao@fudan.edu.cn}
\affiliation[5]{organization={Institute of Big Data (IBD), Fudan University},
    city={Shanghai},
    postcode={200433}, 
    country={China}}
\affiliation[6]{organization={National Institute of Intelligent Evaluation and Governance, Fudan University},
    city={Shanghai},
    postcode={200433}, 
    country={China}}
\credit{Conceptualization, Investigation, Writing - review \& editing}

\author[7]{Wei Lu}
\ead{weilu@whu.edu.cn}
\credit{Conceptualization, Investigation, Writing - review \& editing}
\affiliation[7]{organization={School of Information Management, Wuhan University},
    city={Wuhan},
    postcode={430072}, 
    country={China}}

\cortext[cor1]{Corresponding author.}

\begin{abstract}
The potential impact of a paper is often quantified by how many citations it will receive. However, most commonly used models may underestimate the influence of newly published papers over time, and fail to encapsulate this dynamics of citation network into the graph. In this study, we construct hierarchical and heterogeneous graphs for target papers with an annual perspective. The constructed graphs can record the annual dynamics of target papers' scientific context information. Then, a novel graph neural network, \textbf{H}ierarchical and \textbf{H}eterogeneous \textbf{C}ontrastive \textbf{G}raph \textbf{L}earning Model (\textbf{H$^2$CGL}), is proposed to incorporate heterogeneity and dynamics of the citation network. H$^2$CGL separately aggregates the heterogeneous information for each year and prioritizes the highly-cited papers and relationships among references, citations, and the target paper. It then employs a weighted GIN to capture dynamics between heterogeneous subgraphs over years. Moreover, it leverages contrastive learning to make the graph representations more sensitive to potential citations. In particular, co-cited or co-citing papers of the target paper with large citation gaps are taken as hard negative samples, while randomly dropping low-cited papers could generate positive samples. Extensive experimental results on two scholarly datasets demonstrate that the proposed H$^2$CGL significantly outperforms a series of baseline approaches for both previously and freshly published papers. Additional analyses highlight the significance of the proposed modules. Our codes and settings have been released on Github. (\url{https://github.com/ECNU-Text-Computing/H2CGL})
\end{abstract}

\begin{highlights}
    
    \item We construct hierarchical and heterogeneous graphs to record dynamics of the citation network .
    
    \item We propose a novel model H$^2$CGL to aggregate structural and temporal features.
    
    \item The sensitivity of graph representations to potential citations is improved by contrastive learning.
    
    \item Experimental results show that our model can reasonably make predictions.
    
\end{highlights}


\maketitle

\section{Introduction}
\label{sec:intro}

Assessing the potential impact of papers is of great significance to both academia and industry \citep{wang2013ctask}, especially given the exponential annual growth in the number of papers \citep{lo2020s2orc,chu2021slowed,xue2023dgcbert}. As the numerical value of the scientific impact could be difficult to determine, citation count is frequently employed as a rough estimate \citep{evans2009open, sinatra2016quantifying,jiang2021hints}. Actually, the dynamics in citation networks cannot be ignored. For example, the ``sleeping beauties'' \citep{van2004sleeping} phenomenon indicates that the citations of a paper can vary considerably in different time periods. Besides the content quality, the future citations of a paper will be influenced by newly published papers \citep{funk2017dynamic, park2023papers}. New papers may be successors to older ones, discovering the importance of previous works, thereby drawing more citations for them; or new papers may be competing with older ones, correcting or improving the previous works, thus making them lose potential citations. Therefore, it's imperative to capture dynamics of the citation network to accurately predict the future citations of a target paper.

\begin{figure*}[h]
  \centering
  \includegraphics[width=0.9\textwidth]{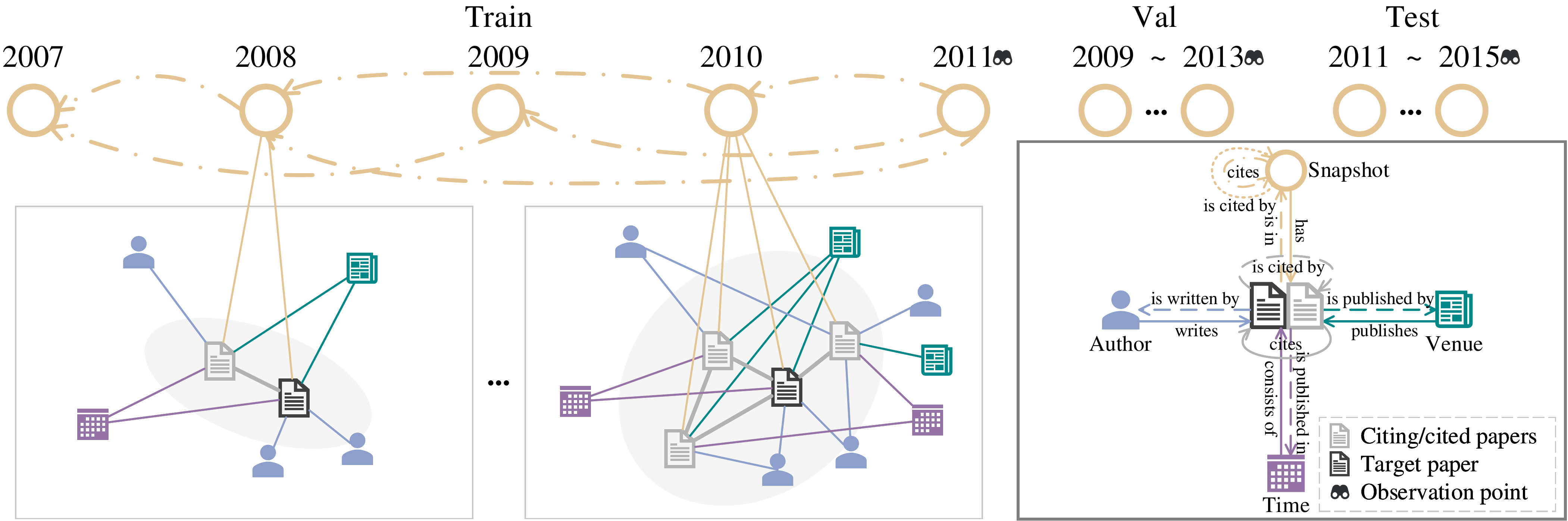}
  \caption{An example of constructing a hierarchical and heterogeneous graph from a citation network for a target paper. The observation points of the paper in the training, validation, and test set are 2011, 2013, and 2015, respectively. The potential citation counts are calculated in 5 years later, \textit{i.e.}, 2016, 2018, and 2020, respectively. For each year before the observation point, a heterogeneous subgraph consists of papers surrounding the target paper, their metadata, and a virtual node snapshot. The snapshot is connected with all papers. We keep snapshots from the past 5 years (if any). They are connected to each other according to whether there are citations among their associated papers, forming the hierarchical graph.}
  \label{fig:sample}
\end{figure*}

Previous studies within informetrics have primarily concentrated on content information or citation networks of papers. Content-based models use Word2Vec \citep{mikolov2013distributed} or BERT \citep{devlin2018bert} for text representation, followed by the use of recurrent neural networks (RNN) or self-attention layers to encode paragraphs or citations \citep{ma2021deep, van2020schubert}. These models incorporate various information, including the content, metadata, and citation sequences of target papers, to predict paper impact based on static features. However, they may not fully exploit the temporal and structural relationships presented in citation networks. Citation network-based models benefit from advances in graph embeddings \citep{perozzi2014deepwalk} and graph neural networks \citep{scarselli2008graph}. Current models, such as MUCas \citep{chen2022mucas}, treat citation as an information diffusion process and view citation predictions as similar to cascade predictions in social networks \citep{li2017deepcas,cao2017deephawkes}. In detail, the citation cascades are extracted with citing and cited papers. Then, they always employ RNNs to encode the sequence of cascades. However, it's important to note that citation networks differ from social networks. For example, the latter can be seen naturally as a tree: a post can only ``cite'' one post while a paper can cite multiple papers. Recently, HINTS \citep{jiang2021hints} as a dynamic GNN gathers metadata of papers and encodes their relations and historical trends via the Heterogeneous Information Network (HIN) followed by the sequence model. It leverages metadata to enable the prediction for new papers that lack accumulated citations. Nevertheless, this model overlooks predictions for previous papers, which are also influenced by new papers. In summary, sequence models such as RNN could fail to accurately represent dynamics in the citation networks. They also neglect the possible confusion among neighboring papers in the network.

In summary, content-based methods primarily focus on individual paper features, while graph-based methods overly emphasize the interaction relationships between entities. Consequently, existing methods often rely on just one aspect of citation networks, missing the intricate dynamics within the citation context. Moreover, it is crucial to underscore the unique characteristics of citation networks, for effectively extracting evolving succession and development information over time. We argue that evaluating paper impact should encompass both the papers themselves and the evolving research context simultaneously, adopting an informetrics perspective.

To this end, we derive hierarchical and heterogeneous graphs for target papers from the citation network, so as to incorporate dynamics into the graph. As depicted in Figure \ref{fig:sample}, a paper's heterogeneous subgraphs will expand over the years as influenced by further citations. Particularly, new citations will not only connect to the target paper but also to its references and citations. As previously mentioned, any updates to the structured data can impact potential citations \citep{park2023papers}. Besides, within each heterogeneous subgraph, we include authors, publication time, and published venue as additional nodes to link with papers. These nodes are only incremented with new citations of the target paper in the subsequent years. During initialization: the representation of \textit{paper} node is learned by language model; the publication \textit{time} node summarizes the information of all papers published in that year; the \textit{author} node encapsulates all papers published by this author before this year; the \textit{venue} node gathers all papers published on this venue before this year. In this way, we can track the dynamics of the scientific community even if the target paper has not been cited in recent years. For each year, we introduce a virtual node, \textit{snapshot}, to aggregate all papers within the heterogeneous subgraph. Further, the yearly changes for each paper are discontinuous and complex \citep{van2004sleeping}. In this approach, snapshots can be linked to each other based on the presence of citations among their associated papers, rather than being modeled sequentially. Thus, the constructed graph can effectively preserve detailed structures in space and enhance complex interactions between snapshots in time. Finally, we predict potential citation counts based on the constructed graphs. Besides, training, validation, and test sets are collected in terms of corresponding observation points. Compared with randomly dividing the dataset regardless of temporal information, this setting is more consistent with the actual usage scenarios.

Based on the proposed graph, we devise a novel end-to-end model named \textbf{H}ierarchical and \textbf{H}eterogeneous \textbf{C}ontrastive \textbf{G}raph \textbf{L}earning model (\textbf{H$^2$CGL}) to predict potential citation counts for target papers. The GNN encoder of H$^2$CGL aggregates different relations individually, similar to relational graph convolutional networks (R-GCN). It consists of multiple layers with two alternating components to encode the structural relations first and then the temporal relation. In the first part, we propose Citation-aware Graph Isomorphism Network (C-GIN) to enhance the importance of global highly-cited papers in the citing relations. Then, Relation-aware Graph Attention Network (R-GAT) is proposed to re-weight the importance of references, citations, and the target paper, which is assisted with the interval from publication to snapshot time. Other relations in each heterogeneous subgraph are encoded by general GINs. The second part is a weighted GIN to figure out the dynamics among yearly snapshots. The edges are weighted by the number of citations between snapshots. The final graph embedding of the target paper is aggregated by sum-pooling of snapshots. Moreover, in the training strategy, we employ contrastive learning adapted to the citation networks to make the graph representations of globally-neighboring papers more sensitive to the potential citation. Specifically, the positive samples are generated via randomly dropping lowly-cited papers. More importantly, We select hard negative samples from the co-cited or co-citing papers with different potential citations. Eventually, the final representation is fed into two-layer MLPs to produce predictions. In this study, besides the potential citation count prediction, we divide citation counts into three intervals (\textit{i.e.}, $< 10$, $10-100$, $\ge 100$) and conduct the interval classification jointly. 

Extensive experiments are conducted on two real-world scientific datasets (PMC and DBLP). The results demonstrate that H$^2$CGL outperforms all alternative models for both previous and fresh papers. Ablation tests demonstrate the importance of the proposed components. Additional results provide insights into the task and model. 

To sum up, the contributions of this research are as follows: 

$\bullet$ To predict potential citation counts for papers, we construct hierarchical and heterogeneous graphs from the citation network. They enable tracking dynamics of previously and freshly published papers in the graph.

$\bullet$ We propose a novel model called Hierarchical and Heterogeneous Contrastive Graph Learning (H$^2$CGL). Graph representations of target papers are aggregated by combining structural and temporal features influenced by new papers. Their sensitivity to potential citations is further improved by contrastive learning.

$\bullet$ Experimental results demonstrate that our model is superior to previous models in terms of both previous papers and fresh papers. Further analyses illustrate that our model can reasonably make predictions.

\section{Theoretical and practical implications}

From a theoretical point of view, our work offers an innovative methodology for estimating the impact of academic papers. First, instead of using traditional static graph sequences, we suggest employing hierarchical and heterogeneous graphs, to better capture the dynamic and structural characteristics of the citation network. Second, we propose a simple yet effective contrastive learning method tailored for citation networks. By comparing papers with negative samples of similar topics collected from co-citing or co-cited papers, the model will be more sensitive to citation dynamics rather than topic diversity. Moreover, we evaluate the proposed model based on chronologically partitioned data instead of randomly partitioned data. This setup is more in line with real-world challenges. This setup accounts for the requirement to handle both newly published papers and predict the future potential of papers from the past. To the best of our knowledge, this is the first work that actually examines model generalization in light of both previous and new papers.

From a practical point of view, our work can bring positive effects to academic evaluation. As the volume of papers grows, the process of knowledge transfer becomes increasingly complex, posing challenges in estimating the impact of numerous papers. Our proposed model facilitates the evaluation of both previous and newly published papers by leveraging comprehensive features extracted from hierarchical and heterogeneous graphs. Thus, policy makers and funders could assess the potential impact of a paper within the existing research landscape. This study complements the current assessment process with an automated approach, enhancing the efficiency and accuracy of academic evaluation.

\section{Related Work}
\label{sec:relate}

We review three lines of related work: citation count prediction, cascade prediction, and dynamic graph neural network. 

\subsection{Citation Count Prediction}

Citation count prediction enables potential impact estimation, which is an important sub-task of automatic academic evaluation. It relies heavily on comprehensive feature extraction from academic entities. Many models for citation prediction have been proposed in bibliometrics, including stochastic models, feature-based models, and deep learning models. Since the distribution of long-term citation counts follows Zipf-Mandelbrot's law \citep{silagadze1997citations}, stochastic models can predict future citation counts by fitting the curve of citation counts \citep{glanzel1995predictive}. Recently, machine learning methods, such as support vector regression and CART, have yielded meaningful results by utilizing manually derived features from papers, authors, journals, \textit{etc} \citep{yan2011citation,yu2014citation,ruan2020predicting}. 
Deep learning techniques now predominate in this challenge benefiting from the quick development of deep neural networks. These models can leverage the success of natural language processing and computer vision to extract more abstract features from the content of papers \citep{abrishami2019predicting,huang2022fine,xue2023dgcbert}. However, most of the current methods tend to focus solely on the features of individual papers or common features shared among groups, while neglecting the relationships between papers and their connections with other entities. In other words, they do not fully exploit the valuable information contained in the citation networks.

Therefore, various models have been developed to utilize citation network information, such as graph embedding methods and Graph Neural Network (GNN) models. SI-HDGNN \citep{xu2022hdgnn} first employs DeepWalk to initialize node embeddings of a heterogeneous graph and then applies Bi-RNN and multi-head attention to aggregate information by types. After getting the final node embeddings, it uses Bi-RNN to encode different meta nodes of citing papers in sequence and get the final prediction through a Multi-Layer Perceptron (MLP). Nonetheless, due to the limitations of graph embedding techniques and static problem settings, SI-HDGNN can only predict papers that are presented in the training dataset and is incapable of handling papers that are published in new years. HINTS \citep{jiang2021hints} applies GCN on pseudo meta-data graphs for cold starting problems and predicts with stochastic models. It combines meta-data information including authors, venues, and keywords, and utilizes GRU to encode temporal information. However, it just models limited graphs of different times separately, and randomizes the features of the nodes, neglecting the rich semantic information contained in the paper content. Moreover, HINTS focuses only on newly-published papers in the predicted years, disregarding the potential of previously-published papers.

In conclusion, existing citation count prediction models tend to focus on one limited aspect of citation networks. In contrast, our approach aims to combine both the features of individual papers and their interaction relationships simultaneously for a more comprehensive analysis. Moreover, we argue that the current evaluation methods are incompatible with practical situations. Since we want to predict the potential impact of papers in the future with current data, we should not just split the data into training/validation/test sets straightforwardly with all available data. Instead, we should use the data before a specific time (like 2011) as the training set and try to predict all the papers in the future at another specific time (like 2015). The test data should contain both previous papers in the training sets and fresh papers published during the interval. We require a model that can not only forecast the increased citations of existing papers but also accommodate newly-published ones.

\subsection{Cascade Prediction}
Cascade prediction has been a long-standing and critical problem in the field of information diffusion and social network analysis \citep{zhou2021cascade}. The primary goal of cascade prediction is to anticipate the popularity of a post, which can involve either a regression task, where the exact number is predicted, or a link prediction task, which predicts the users who might interact with the post. In the context of citation networks, citation prediction can also be regarded as a form of cascade prediction. In this case, the paper serves as the post, and the authors or author groups serve as the users of the social network. Similar to citation prediction, proposed models for cascade prediction can be classified into stochastic models, feature-based models, and deep learning models.

Stochastic models take them as time-series data and adopt various stochastic processes to model and simulate the information diffusion in networks, in a statistical and generative manner \citep{zhao2015seismic, gao2020p-cascade}. Feature-based models also extract manual features and apply machine learning methods to predict the cascade \citep{guille2013cfeature, tatar2014cfeature2}. Recently, deep neural models have also dominated the cascade prediction tasks. They extract structural information through graph embedding, sequence models, GNN models, and so on, and then encode the temporal information with sequence models like RNN. DeepCas \citep{li2017deepcas} utilizes DeepWalk for graph embeddings and feeds them into RNNs for cascade modeling and prediction. DeepHawkes \citep{cao2017deephawkes} employs Hawkes points process and RNN models for cascade prediction. GTGCN \citep{yang2022gtgcn} utilizes the temporal embedding and integrates it into BiGCN and GRU. MUCas \citep{chen2022mucas} combines the BiGCN and capsule networks to encode bidirectional and high-order information. CCGL \citep{xu2022ccgl} uses contrastive learning with augmentation methods designed for cascade graphs and distills it to obtain the final model.

However, we should realize that citation networks have some unique characteristics distinct from social networks. For instance, the act of citing differs from that of retweeting, as papers only have immediate successors and lack strict mediation to propagate information. In contrast, the cascade graph naturally represents a tree-like structure for information diffusion. Citing behavior is much more intricate than retweeting. Although we can transform the citation network to the citation cascade to imitate the structure of the cascade graph, this transformation may result in the loss of valuable information. Moreover, the accumulation of citations takes much longer than the spread of posts on social networks. Some papers may have few or even no citations for a long time, making reliance solely on the paper's cascade graph limited in addressing this issue. To tackle this challenge, we propose constructing target-centric graphs of the target paper that incorporate both references and citations. We also leverage heterogeneous graphs with authors and journals to integrate global information from other entities. This approach allows us to partially overcome the cold start problem in citation/cascade prediction by exploiting the rich information contained within these graphs.

\subsection{Dynamic and Heterogeneous Graph Neural Networks}

Graph Neural Networks \citep{sperduti1997supervised,gori2005new,scarselli2008graph,wu2020comprehensive} are widely applied to deal with non-Euclidean data, such as graphs with complex relationships and interdependency between nodes. Dynamic graphs are characterized by their temporal nature, reflecting changes and evolutions over time, while heterogeneous graphs involve multiple node types or edge types. The inclusion of these temporal and multi-dimensional aspects makes dynamic and heterogeneous graphs significantly more complex than static homogeneous graphs. Therefore, analyzing and utilizing dynamic and heterogeneous graphs require the development and application of diverse and advanced methodologies. In this section, we conduct a comprehensive review of recent advancements in the field of Dynamic GNNs and Heterogeneous GNNs, separately examining these topics to provide a clear and focused analysis.

For dynamic graphs, they can be divided into continuous and discrete dynamic graphs. Continuous graphs are dynamic graphs that vary almost at all times, while discrete graphs change after a period of time and can be directly divided into many snapshots. Currently, studies about dynamic graph neural networks mainly focus on discrete dynamic graphs. Many methods first utilize static GCN as the structural encoder and then utilize sequential models such as RNNs and transformers to process the obtained embedding sequence, including DGCN\citep{manessi2020dgcn} and Dysat\citep{sankar2020dysat}. Also, some models integrate GCN into RNN models by setting the parameters in RNN cells as GCN parameters \citep{pareja2020evolvegcn}. Recently, ROLAND \citep{you2022roland} is proposed to challenge the effectiveness of GCN equipped with sequential models. The researchers argue that such models overlook the significance of different orders of information. To address this limitation, ROLAND first uses the output of all layers of GCN, including edge features, and then employs an updater like GRU to encode hierarchical temporal information. For continuous dynamic graphs, researchers commonly transform them into discrete dynamic graphs with multiple snapshots, or add time-stamp information in a single static graph. Then, they apply discrete methods with some adaptation to the continuous ones. For example, Graph2Route \citep{wen2022graph2route} just divides the continuous take-and-delivery into different snapshots and then uses temporal-correlation encoding to extract temporal information. Also, there are some works designed directly for continuous ones. For instance, InstantGNN \citep{zheng2022continuousgnn} extends the basic propagation method for edge and attribute changes.

The separate message-passing method, exemplified by Relational Graph Convolutional Networks (R-GCN) \citep{schlichtkrull2018rgcn}, first performs message passing within individual relations, followed by aggregation of information from these relations to update node representations. This approach facilitates targeted information to flow across different types of relationships within the heterogeneous graph. In contrast, the relation-learning method transforms the heterogeneous graph into a homogeneous graph while preserving the node and edge type indicators to retain heterogeneous information. An example of this approach is Heterogeneous Graph Transformer (HGT) \citep{hu2020hgt}, which utilizes multi-head attention modules with specific parameters tailored to different node and edge types, effectively modeling the diverse nature of the heterogeneous graph. ROLAND \citep{you2022roland}, also encodes heterogeneous graphs and addresses the challenge by projecting the edge types onto distinct edge embeddings. This proves instrumental in effectively representing and capturing the heterogeneous information within the graph structure.

Specially, there are two challenges to encoding the dynamic citation networks: 1) How to represent the complex relationships between different snapshots? While several methods have been proposed to handle dynamic graphs, many of them simply combine GNN models and sequential models without taking into account the intricate and detailed information between different time points. Instead, we argue that the temporal information can also be represented as graphs, so we can form the dynamic graphs as hierarchical structures to capture this information more effectively. This approach facilitates a better understanding of the underlying dynamics. By encoding both the temporal and structural information of dynamic graphs, we can capture detailed information across different snapshots and represent potential interactions between them. 2) How to cope with the distinct characteristics of citation networks? Citation graph is a unique type of discrete graph, where new edges and nodes are continuously added over time and do not disappear, making it inherently dynamic. After new edges and nodes appear, they will not disappear and become static. Most current methods cannot deal with the changes in the number of nodes and cannot model the small variation of edges between different snapshots in citation networks. Moreover, citation graphs can be extremely large, with a high number of nodes and edges, making it challenging to encode the entire graph. Therefore, we typically only obtain subgraphs of predicted papers. To better model the snapshots, we choose R-GCN as the backbone of our proposed model, as it offers the ability to model separate relations more attentively and can leverage customized variants based on the SOTA GCNs, providing further benefits. More specifically, we design the C-GIN to model the difference of citation and R-GAT for snapshot nodes to gather the information from different types of papers (reference, target, and citation papers). To better exploit the difference, we apply contrastive learning with data augmentation and negative sampling specifically designed for the citation network. By selecting the papers with related topics as hard negative samples, the model can differentiate similar papers and encode them more precisely.
\section{Methodology}
\label{sec:method}

\begin{figure*}[!h]
  \centering
  \includegraphics[width=0.97\textwidth]{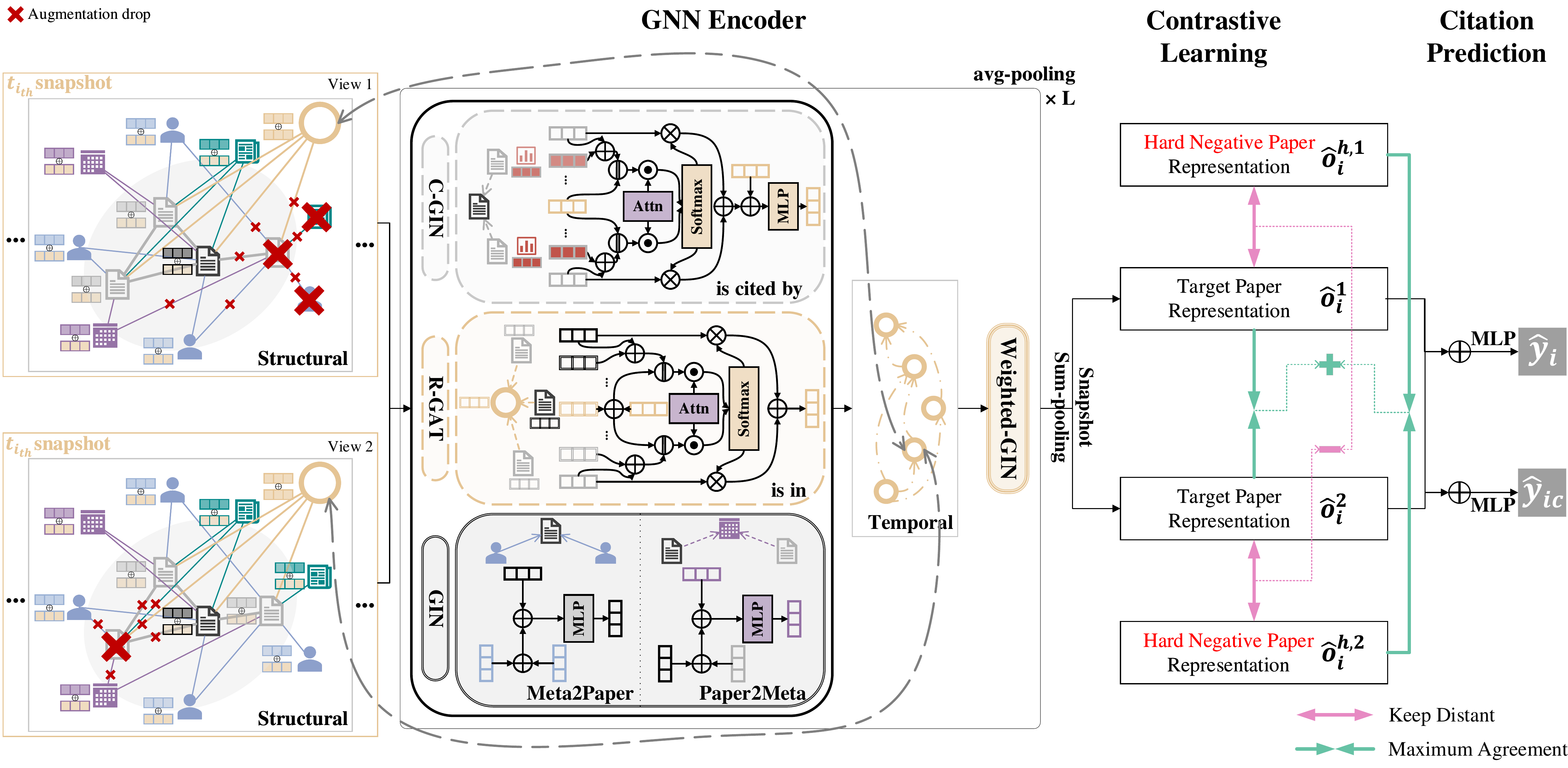}
  \caption{The overall architecture of H$^2$CGL. The hierarchical and heterogeneous graphs of target papers are encoded by the GNN encoder. It first utilizes the shared customized R-GCN to encode the \textbf{structural} information within subgraphs at different snapshot times simultaneously. Specially, C-GIN is designed for \textit{paper} nodes in ``is cited by'' edges to focus on highly-cited ones, and R-GAT is designed for \textit{snapshot} nodes in ``is in'' edges to aggregate the \textit{paper} nodes in subgraphs based on their types and target paper publication age. Next, the weighted-GIN is applied to extract the complex \textbf{temporal} information within the snapshot graphs. Then the target paper representations are obtained through sum-pooling across the time dimension. After that, the representations are further improved by a contrastive learning module. In the training phase, it drops the \textit{paper} nodes within different subgraphs to extract the truck of knowledge diffusion and samples co-cited or co-citing papers as the hard-negative candidates. Finally, they are fed into two MLPs to jointly predict potential citation counts and intervals in the future.}
\label{fig:model}
\end{figure*}

In this section, we introduce our proposed model Hierarchical and Heterogeneous Contrastive Graph Learning Model (\textbf{H$^2$CGL}) in detail. The overall framework of H$^2$CGL is depicted in Figure \ref{fig:model}. 

\subsection{Problem Formulation and Notations}

Before presenting our approach, we first introduce the problem formulation and notations. In reality, the citation network consists of many heterogeneous entities and relations, which are dynamically evolving each year. To simulate the entity-aware and time-variant information, we construct hierarchical and heterogeneous graphs for target papers from the citation network to predict potential citation counts.

\subsubsection{Hierarchical and Heterogeneous Graph}
\label{sec:h2g}

Given a target paper at the time step $t$, its heterogeneous subgraph is defined as $G^t = (\nu^t, \epsilon^t, \phi, \varphi)$. Here, $\nu^t$ denotes the set of nodes surrounding the target paper at the time step $t$. And $\epsilon^t$ represents the set of edges among $\nu^t$. Each node $\nu$ and edge $\epsilon$ are associated with their type mapping functions $\phi:\nu \rightarrow U$ and $\varphi:\epsilon \rightarrow R$, where $U$ and $R$ denote the types of nodes and edges, respectively (see Figure \ref{fig:sample}). 

Specifically, the backbone of the heterogeneous subgraph is constructed by \textit{citing} and \textit{cited} relations. We collect references and 1-hop citations of the target paper at the time $t$. Additionally, links exist between the reference papers and citation papers. Since some target papers have plenty of references, we just maintain the top $K$ newest published and most cited ones. The same procedure is used to select citations. This can help to save the most recent and important information. Metadata nodes associated with papers are introduced into the subgraph to complement the lack of global information. We further introduce a virtual node \textit{snapshot} into the subgraph and link it to all papers. It will be employed to aggregate the structural information of the target paper at the time $t$ (see Figure \ref{fig:sample}). 

Before the observation point, there will be $T$ subgraphs of the target paper at all time steps. Then, the proposed hierarchical graph of the target paper is established on the set of $T$ heterogeneous subgraphs. It is denoted as $G=(<G^t>_{t=1}^T, \epsilon)$, where $<G^t>_{t=1}^T=\{G^1, \cdots, G^t, \cdots, G^T\}$ and $\epsilon$ denotes edges between different subgraphs. These edges are directly linked between $T$ \textit{snapshot} nodes and defined by whether there are citations between their associated papers. Here, we can take heterogeneous subgraph at each time step as the low-level data with structural information. The graph of subgraphs at all time steps serves as the high-level data with temporal information.

\subsubsection{Scientific Impact Prediction}

In this study, we estimate the scientific impact of a target paper by predicting its potential citation count. It is defined as the increased citation number $\Delta$ years after the observation point. We also divide the increased number into three intervals: $<10$, $10-100$, and $\ge 100$. Given a paper, we try to learn a function $f(\cdot)$ to jointly predict the citation count and estimate the probability of the citation interval.

\subsection{Graph Neural Network Encoder}

Here, we describe how we model the proposed hierarchical and heterogeneous graphs via graph learning. The encoder includes node embedding initialization and modified relational GCN. We make the hierarchical model part of the GNN.

\subsubsection{Initialization}

The initial embeddings of \textit{paper} nodes are the text embeddings extracted from the combination of the title and abstract. Embeddings of the metadata nodes are averaged by all papers associated with them in the global citation network at that time. These papers can be beyond the subgraph of the target paper (also described in Section \ref{sec:intro}). The text embedding can be extracted by the averaged Word2Vec, GloVe, or BERT embeddings. Particularly, we initialize the embedding of the \textit{snapshot} node by the time embedding of that \textit{snapshot} time point. In addition, we also calculate some node features such as the global citation counts of papers for each year. For a node $v$, we have:
\begin{equation}
    \begin{split}
        & \bm{h}^{t, paper}_v = TextEncoder([w_1, w_2, ..., w_{n^w}]) \\
        & \bm{h}^{t, u}_v = avgpool([\bm{h}^{t, paper}_1, \bm{h}^{t, paper}_2, ..., \bm{h}^{t, paper}_{m_p}]), \\
        & u \in \{author, venue, time\}
    \end{split}
\label{eq:text_encoder}
\end{equation} 
where $u$ denotes the node type, the $TextEncoder$ encodes the title and abstract of papers, other node types are extracted from the avg-pooling of the papers they connected, and $m_p$ is the count of its connected papers.

Assume that the general pattern of knowledge flow and scientific diffusion remains relatively constant across years. Thus, the structural characteristics of the subgraphs at different time points have commonalities. Therefore, we plan to encode the information of different time points simultaneously. We only employ a single bidirectional hierarchical relational GCN for all subgraphs, instead of applying different encoders for different time points. To capture dynamics, we need to further emphasize temporal information for each subgraph. Here, we attach the corresponding time point information to the node embeddings before they are fed into the structure encoder. To be specific, we project the time point into embeddings for different types of nodes to model the time context-aware information and add it to the original node embeddings $\bm{h}^{t, u}_v$. Therefore, we can get the new node embeddings $\bm{z}^{t, u}_v$:
\begin{equation}
    \begin{split}
       & \bm{z}^{t, u}_v = \bm{h}^{t, u}_v + f_{ste}(u, t), u \in U
    \end{split}
\label{eq:ste}
\end{equation} 
where $f_{ste}$ is the time point embedding function that map the node type $u$ and time point $t$ into embedding.

\subsubsection{Modified Relational GCN}

The heterogeneous graphs have multiple node types and edge types. Generally, we apply Relational GCN (R-GCN) \citep{schlichtkrull2018rgcn} to model distinct information according to edge types with a hierarchical manner. As we want to encode the detailed information of hierarchical graphs, we divide the R-GCN for modeling all edges and node types simultaneously into a structural encoder and a temporal encoder. In the structural encoder, we only conduct information passing and node aggregation inside heterogeneous subgraphs to model the structural information. To aggregate information from different edges, we sum them up to get the aggregated embeddings. The \textit{snapshot} nodes are updated by papers they linked, and the other types of nodes can be only updated by the nodes in the same heterogeneous subgraph.

After we have updated all nodes inside the heterogeneous subgraphs separately, we conduct GCN only on \textit{snapshot} nodes to model the temporal information. We stack $l$ layers to encode multi-hop and high-order information in structural and temporal dimensions alternatively. Compared to the old methods that first apply GCN for subgraphs separately and then feed the pooled embedding to the sequential model, we can encode more complex joint interactions of structural and temporal features.

We briefly use $\bm{z}^{(l)}$ to represent specified hidden representation of different types in the $l_{th}$ layer and $\bm{o}$ to be the final output embedding of the target paper aggregated from \textit{snapshot} nodes:
\begin{equation}
    \begin{split}
        & \bm{z}_{dst}^{(l+1)} = \underset{r\in\mathcal{R}^{intra}, r_{dst}=dst}{AGG} 
        (f_r(g_r, \bm{z}_{r_{src}}^{(l)}, \bm{z}_{r_{dst}}^{(l)})) \\
        & \hat{\bm{z}}_{dst}^{(l+1)} = \underset{r\in\mathcal{R}^{inter}, r_{dst}=dst}{AGG} (f_r(g_r, \bm{z}_{r_{src}}^{(l+1)}, \bm{z}_{r_{dst}}^{(l+1)})) \\
        & \Tilde{\hat{\bm{\bm{Z}}}} = avgpool([\bm{W}^{(1)}\hat{\bm{Z}}^{(1)}, \bm{W}^{(2)}\hat{\bm{Z}}^{(2)}, ..., \bm{W}^{(L)}\hat{\bm{Z}}^{(L)}])\\
        & \bm{o} = sumpool(\Tilde{\hat{\bm{z}}}^{s})
    \end{split}
\label{eq:graph_encoder}
\end{equation} 
where $\mathcal{R}^{intra}$ is the set contains all edges inside heterogeneous subgraphs, $\mathcal{R}^{inter}$ is the set contains all edges between heterogeneous subgraphs, $f_r$ is the neural network module for each relation $r$, and $AGG$ is the relation aggregation function in the structural encoder.

Instead of just applying classic GCN in different edge types, we replace and modify the powerful graph neural network model for capturing citing-aware information to better model the scientific diffusion of citation. Specifically, we design Citation-aware GIN (C-GIN) to enhance the importance of highly-cited papers in citing behavior. Then we apply Relation-aware GAT (R-GAT) for \textit{snapshot} nodes to integrate information of references, citations, and target papers according to the time after publication. For other edges, we just apply the common GIN \citep{xu2018gin} in sum pooling to retain the representation power close to 1-WL test:
\begin{equation}
    \begin{split}
         \bm{z}_{i, r_{dst}}^{(l+1)} = f_{\Theta, r} ((1 + \xi) \bm{z}_{i, r_{dst}}^{(l)} + AGG(\{\bm{z}_{j, r_{src}}^{(l)}, &j\in\mathcal{N}(i, r_{src})\})), r\in\mathcal{R}^{normal}
    \end{split}
\label{eq:gin}
\end{equation} 
where $\xi$ is the hyper-parameter to decide the retained original information, $\mathcal{R}^{normal}$ is the set including all edges in classic GIN, and $AGG$ is the neighborhood aggregation function of the GIN.

\subsubsection{Citation-aware GIN}

C-GIN replaces the simple sum pooling of GIN with an attention mechanism based on the global citation information of citing papers. It is only applied on the ``cites'' edges. If a paper is cited by a highly-cited paper, the potential citations that the paper may receive in the future will also increase further. To some extent, it can demonstrate the value of this paper. In contrast, whether a paper cites highly-cited papers is of limited use, as these papers are often cited by many average papers as well. Notably, this paper may be influenced by the highly-cited papers, but that does not mean this paper has equal value as the highly-cited paper. Hence, we apply C-GIN on the ``cites'' edges. Here, we apply the positional embedding \citep{vaswani2017attention} as the degree embedding since it can represent the degree and keep the same order of magnitude.  
\begin{equation}
    \begin{split}
         &\bm{z}_{i, r_{dst}}^{(l+1)} =f_{\Theta, r} ((1 + \xi) \bm{z}_{i, r_{dst}}^{(l)}\\
         &\qquad\quad + AGG(\{\alpha_{ji}^{(l)} \bm{z}_{j, r_{src}}^{(l)}, j\in\mathcal{N}(i, r_{src})\}))\\
         &\alpha_{ij}^{(l)} =\mathrm{softmax_i} (e_{ij}^{(l)})\\
         &e_{ji}^{(l)} ={\bm{W}_a^{(l)}}\mathrm{LeakyReLU}(f^{(l)}_{ce}(n_{r_{src}}^c) + mlp^{(l)}_{src}(\bm{z}_{j, r_{src}}^{(l)}) + mlp^{(l)}_{dst}(\bm{z}_{i, r_{dst}}^{(l)}))
    \end{split}
\label{eq:cgin}
\end{equation} 
where $f^{(l)}_{ce}$ is the citation degree embedding, $mlp^{(l)}_{src}$ and $ mlp^{(l)}_{dst}$ are MLPs that transform the source and target nodes, $\bm{W}_a^{(l)}$ is the attention matrix, and $n_{r_{src}}^c$ is the citation of the source paper.

\subsubsection{Relation-aware GAT}
R-GAT modifies the attention mechanism of GATv2 \citep{brody2021gatv2} to better aggregate the papers inside heterogeneous subgraphs. We assume that for papers published at different times, the references, citations, and target papers may be of different importance. For example, for newly-published papers, they have not accumulated enough citations, so the reference papers and target paper itself may have more significant information to estimate the potential citation. In contrast, for some old papers, the newly-appeared citation papers may be more important to represent the novel breakthrough in recent years, which can be similar to ``sleeping beauties'' phenomenon. Therefore, we categorize the \textit{snapshot} nodes into $c$ groups based on the snapshot time after the publication time of target papers. Also, we divide the \textit{paper} nodes inside subgraphs into reference, citation, and target papers. We use embedding layer to project these categories into different type embeddings, and add them into the attention embedding before getting the attention score.
\begin{equation}
    \begin{split}
        & \bm{z}_{i, s}^{(l+1)} = \sum_{j\in \mathcal{N}(i)} \alpha_{ij}^{(l)} \bm{W}^{(l)}_{right} \bm{z}_{j, paper}^{(l)} \\
        & \alpha_{ij}^{(l)} = \mathrm{softmax_i} (e_{ij}^{(l)})\\
        & e_{ij}^{(l)} = {\bm{W}_a}^{(l)}\mathrm{LeakyReLU}(\bm{W}^{(l)}_{left} \bm{z}_{i, s} + \bm{W}^{(l)}_{right} \bm{z}_{j, paper} + f^{(l)}_{se}(c^{st}_{i}) + f^{(l)}_{pe}(c^{pt}_{j}))
    \end{split}
\label{eq:rgat}
\end{equation} 
where $f^{(l)}_{se}$ is the snapshot type embedding, $f^{(l)}_{pe}$ is the paper type embedding, $c^{st}_{i}$ is the snapshot type of the target snapshot, $c^{pt}_{j}$ is the paper type of the source papers (reference, citation, and target), and $\bm{W}_a^{(l)}$ is the attention matrix.

\subsubsection{Weighted GIN}

Specially, we re-weight the edge weight according to the citation intensity for relations between \textit{snapshot} nodes. Here, we use graph normalization to get the edge weight:
\begin{equation}
    \begin{split}
        & \hat{\bm{z}}_i^{(l+1)} = f_\Theta ((1 + \epsilon) \bm{z}_i^{(l+1)} + AGG(\{e_{ji} \bm{z}_j^{(l+1)}, j\in\mathcal{N}(i)\}))\\
        & e_{ji} = \mathbf{\Tilde{\hat{\bm{A}}}}_{ji} = \mathbf{\bm{D}}^{-1/2} \mathbf{\hat{\bm{A}}}\mathbf{\bm{D}}^{-1/2}
    \end{split}
\label{eq:wgin}
\end{equation} 

We aggregate the neighbors of the target snapshots through weighted edges instead of the sum pooling. The stronger the citation intensity between two snapshots, the more information between the two snapshots is passing. It enables our model to better simulate dynamics across time points.

\subsection{Contrastive Learning Module}
Since we use limited information from the citation network, it may be difficult for the graph encoder to distinguish the potentials between neighboring papers. Therefore, we apply contrastive learning to introduce the self-supervised manner to help the encoder learn distinctly. To elaborate, we follow the augmentation method of CCGL \citep{xu2022ccgl} and make modifications to adapt to the citation network. We retain the method of dropping \textit{paper} nodes according to their degrees, but discard the method of adding nodes. It is widely accepted that highly-cited papers have more influence on other studies, whether the influence may be positive or negative. Therefore, the highly-cited papers are always more important in the process of scientific diffusion. 

Thus, we assign probabilities $p_i$ to \textit{paper} nodes in subgraphs based on the citations received in the global citation network. Apart from the target paper, \textit{paper} nodes with the probability value below $p^d$ are dropped. The metadata nodes of the removed papers will also be dropped if no other papers are connected in the subgraph.
\begin{equation}
    \begin{split}
        & p_i = p^d \cdot n^{paper} \frac{exp(-n_i^c)}{\sum_{j=0}^{n^{paper}}exp(-n_j^c)}
    \end{split}
\end{equation}
where $n^{paper}$ is the \textit{paper} node count of the augmented subgraph and $n_i^c$ is the citation count of paper $i$.

In the citation network study, the relationship of co-reference and co-citation is important, as these papers may have similar study problems or apply equal methods. In other words, papers may be more similar to these papers than other papers with different topics and methods. Here, we sample the co-reference or co-citation papers that possess different citation interval labels from the target paper as hard negative candidates. For each epoch, we sample $n^h$ hard negative samples in the candidate sets with equal probability and feed them into the same encoder of the predicted samples.

For dual-view contrasting, we apply the augmentation methods twice to generate different views. We only contrast the final embedding $\bm{o}$ generated by the encoder to simplify the process. Additionally, we do not contrast each sample in the batch in a pair-wise manner. Instead, we only contrast the embeddings between the predicted paper and its hard negative samples. In this manner, we motivate the model to learn the distinct information consistent with the impact and not related to the topic they study. Therefore, the model can further discriminate breakthrough papers irrelevant to the topic. The topic information will not dominate the model prediction.

The loss function can be formed as:
\begin{equation}
    \begin{split}
        & \hat{\bm{o}}_i = norm(mlp(\bm{o}_i))\\
        & \mathcal{L}_{cl} = \frac {1}{N}\sum_{i=0}^{N}{-log\frac{exp(sim(\hat{\bm{o}}_i^{1}, \hat{\bm{o}}_i^{2})/\tau)}{\sum_{j=0}^{n^h}{(j\neq i)exp(sim(\hat{\bm{o}}_i^{1}, \hat{\bm{o}}_j^{2})/\tau)}}}
    \end{split}
\label{eq:cl_loss}
\end{equation}
where $sim(\cdot)$ is the similarity function like dot product and $\tau$ is temperature parameter.

\subsection{Training}

After getting the final representations from the graph encoder, we feed them into two MLPs.

The main objective of our task is to predict the future increased citation of all samples. Considering that the distribution of citation count is skew distribution, we apply the log transformation to normalize the distribution. 

So, the main loss function of the potential citation count prediction task is MSE between log values:
\begin{equation}
    \begin{split}
        & \hat{y}_i = mlp_r(\bm{o}_i)\\
        & \mathcal{L}_{reg} = \frac {1}{N} \sum_{i=1}^{N}{(y_i - \hat{y}_i)^2}
    \end{split}
\label{eq:reg_loss}
\end{equation}

Additionally, to help better find the breakthrough works of all papers, we add a graph classification task. We split the papers into three intervals according to the increased citation count. We assume that the increased citation count is below 10 as low-cited papers, between 10 and 100 as normal-cited common papers, and over 100 as highly-cited breakthrough papers. Although the labels are extracted from the increased citation in the main regression task, we use original values instead of log values. We motivate the model to represent papers of different categories with distinct representations, to help the regression task. Here we use the cross entropy to calculate the loss:
\begin{equation}
    \begin{split}
        & \hat{y}_{ic} = mlp_c(\bm{o}_i)\\
        & \mathcal{L}_{cls} = \frac{1}{N}\sum_{i} \mathcal{L}_i = - \frac{1}{N}\sum_{i} \sum_{c=1}^My_{ic}\log(\hat{y}_{ic}) 
    \end{split}
\label{eq:cls_loss}
\end{equation}

The final loss function includes the main regression loss, classification loss, and contrastive learning loss:
\begin{equation}
    \begin{split}
        & \mathcal{L} = \mathcal{L}_{reg} + \alpha \mathcal{L}_{cls} + \beta \mathcal{L}_{cl}
    \end{split}
\label{eq:all_loss}    
\end{equation}
where $\alpha$ and $\beta$ are the hyper-parameters to balance the weight of $\mathcal{L}_{cls}$ and $\mathcal{L}_{cl}$.
\section{Experimental Settings}
\label{sec:exp}

In this section, we conduct extensive experiments and analyses on two real-world scientific datasets PMC \citep{pmc2003} and DBLP \citep{tang2008dblp} to validate H$^2$CGL. As a byproduct of this study, we have released the codes and the hyper-parameter settings on the Github to benefit other researchers.\footnote{\url{https://github.com/ECNU-Text-Computing/H2CGL}} Particularly, we want to answer the following research questions (RQ):

$\bullet$ \textbf{RQ1:} Can H$^2$CGL improve the impact prediction task?

$\bullet$ \textbf{RQ2:} What is the role of each component in H$^2$CGL? 

$\bullet$ \textbf{RQ3:} How sensitive is H$^2$CGL to hyper-parameters?

$\bullet$ \textbf{RQ4:} How does H$^2$CGL conduct the prediction?

\subsection{Datasets}

We select widely-used citation network datasets with meta-data information from different fields of study to enhance the generality of our approach. Specifically, we choose PMC \citep{pmc2003} and DBLP \citep{tang2008dblp}, representing medicine and computer science, respectively. These fields are known for their rapid development and exhibit significant differences, making them ideal candidates for our study. Given the two scientific datasets, we construct the corresponding citation networks using the following steps. First, we remove papers without valid meta-information such as author (except for PMC) and venue. We also remove papers with abstracts of less than 20 words. Then, we extract the papers, authors, venues, and publication times as network nodes. They are connected according to the relationships described in Section \ref{sec:method}. Afterwards, we filter out papers published after the test observation point. Finally, for PMC, the whole citation network contains 1,125,788 papers and 9048 venues (\textit{e.g.}, journal). For DBLP, the whole citation network contains 1,786,540 papers, 1,838,509 authors, and 31,665 venues (\textit{e.g.}, journal). The detailed statistics for nodes and edges of the whole citation networks in PMC and DBLP are shown in Table \ref{tab:graph_detail}.

\begin{table*}[htbp]
  \centering
  \caption{Detailed statistics for edges and nodes of the whole citation networks in PMC and DBLP.}
    \begin{tabular}{l|rrr|rrr|r}
    \toprule
    \multicolumn{1}{c|}{\multirow{2}[4]{*}{dataset}} & \multicolumn{3}{c|}{node} & \multicolumn{3}{c|}{edge} & \multicolumn{1}{c}{sparisity} \\
\cmidrule{2-8}          & \multicolumn{1}{c}{paper} & \multicolumn{1}{c}{author} & \multicolumn{1}{c|}{venue} &
\multicolumn{1}{c}{cite} & \multicolumn{1}{c}{write} & \multicolumn{1}{c|}{publish} & \multicolumn{1}{c}{paper} \\
    \midrule
    pubmed (2015) & 1,125,788  & 0     & 9,048  & 3,206,782  & 0     & 1,125,788  & 2.530E-06 \\
    dblp (2013) & 1,786,540  & 1,838,509  & 31,665  & 12,823,956  & 5,007,737  & 1,786,540  & 4.018E-06 \\
    \bottomrule
    \end{tabular}%
  \label{tab:graph_detail}%
\end{table*}%
 
Based on the entire citation network, we locate target papers that our model will address. They should first have more than 5 references. Following MUCas \citep{chen2022mucas}, we have refined our sample selection process to choose papers that offer sufficient and valuable information for model learning. It should be emphasized that we extend its strategy to retain as many samples as possible, enhancing robustness and generality. Instead of limiting samples by their initial size within the observation time window, we now filter papers based on their total accumulated citations since publication after an additional 5 years from the test observation point. In our problem setting, papers are considered if they have more than 10 total citations, even if their initial citations within the observation time window are fewer than 10. After this filtering process, the distribution of objectives becomes closer to a normal distribution. All of these selected papers are included in the test set. Then, we obtain the training set and validation set by filtering papers published after their respective observation points. Thus, the training set is a subset of the validation set, which in turn is a subset of the test set. We use the increased citation count 5 years after the observation point as the ground truth for evaluation. Next, hierarchical and heterogeneous graphs of target papers are extracted from the entire citation network according to corresponding observation points (see Section \ref{sec:h2g}). We retain all papers collected from PMC and randomly sample 120,000 papers from papers collected from DBLP as target papers. Note that since there are few author ids in PMC, no author information is used in this dataset.

It should be noted that even though there are intersections between different sets of papers, their input graphs and objectives (ground truth) may vary significantly due to shifts in the research context. A paper that was considered a hot topic in 2011 may have become less significant by 2015. To maintain a clear distinction between the train and test sets, the observation time for each set is kept at a four-year interval. Additionally, the inclusion of new papers published during this period helps ensure the differences between these sets. In fact, the test set contains a considerably larger number of new papers compared to the training set (see Table \ref{tab:sample_count}). To this end, we can validate the performance of the model for both previous and fresh papers. The observation points of PMC for training, validation, and testing are 2011, 2013, and 2015, respectively. They are 2009, 2011, and 2013 for DBLP. 

\begin{table}[htbp]
  \centering
  \caption{Detailed statistics for training, validation, and testing sets of PMC and DBLP.}
    \begin{tabular}{l|rrrr}
    \toprule
    dataset & \multicolumn{1}{c}{$\textless10$} & \multicolumn{1}{c}{10-100} & \multicolumn{1}{c}{$\geq100$} & \multicolumn{1}{c}{total} \\
    \midrule
    \multicolumn{5}{c}{train} \\
    \midrule
    pubmed & 14,372  & 16,262  & 382   & 31,016  \\
    dblp  & 51,372  & 37,771  & 2,559  & 91,702  \\
    \midrule
    \multicolumn{5}{c}{validation} \\
    \midrule
    pubmed & 27,987  & 37,614  & 903   & 66,504  \\
    dblp  & 64,606  & 40,028  & 3,009  & 107,643  \\
    \midrule
    \multicolumn{5}{c}{test} \\
    \midrule
    pubmed & 41,052  & 83,541  & 2,094  & 126,687  \\
    dblp  & 82,346  & 34,761  & 2,893  & 120,000  \\
    \bottomrule
    \end{tabular}%
  \label{tab:sample_count}%
\end{table}%

\begin{figure*}[]
  \centering
  	\centering
	\subfloat[PMC]{\includegraphics[width=0.22\textwidth]{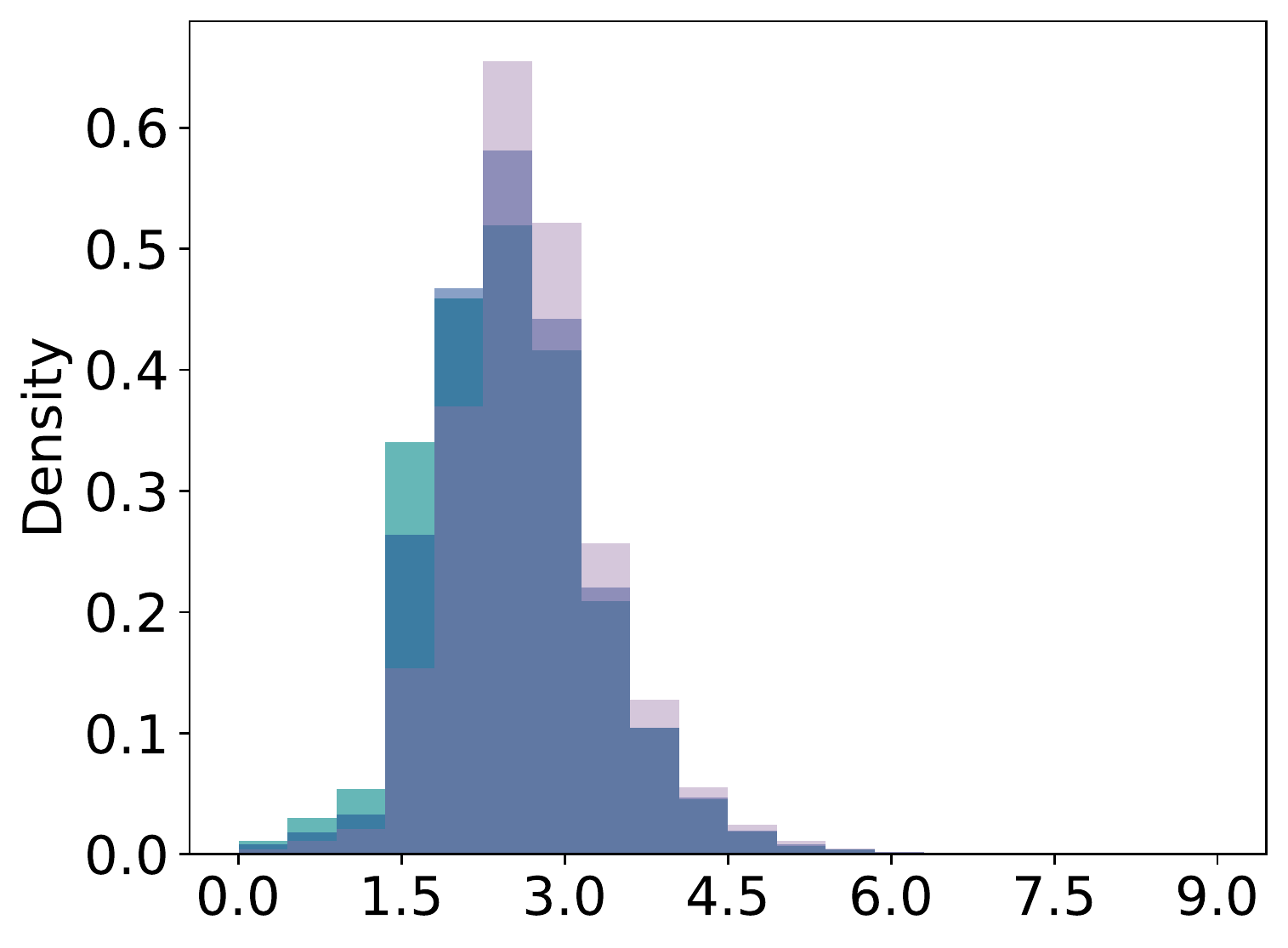}%
		\label{fig:pmc}}
	\subfloat[DBLP]{\includegraphics[width=0.22\textwidth]{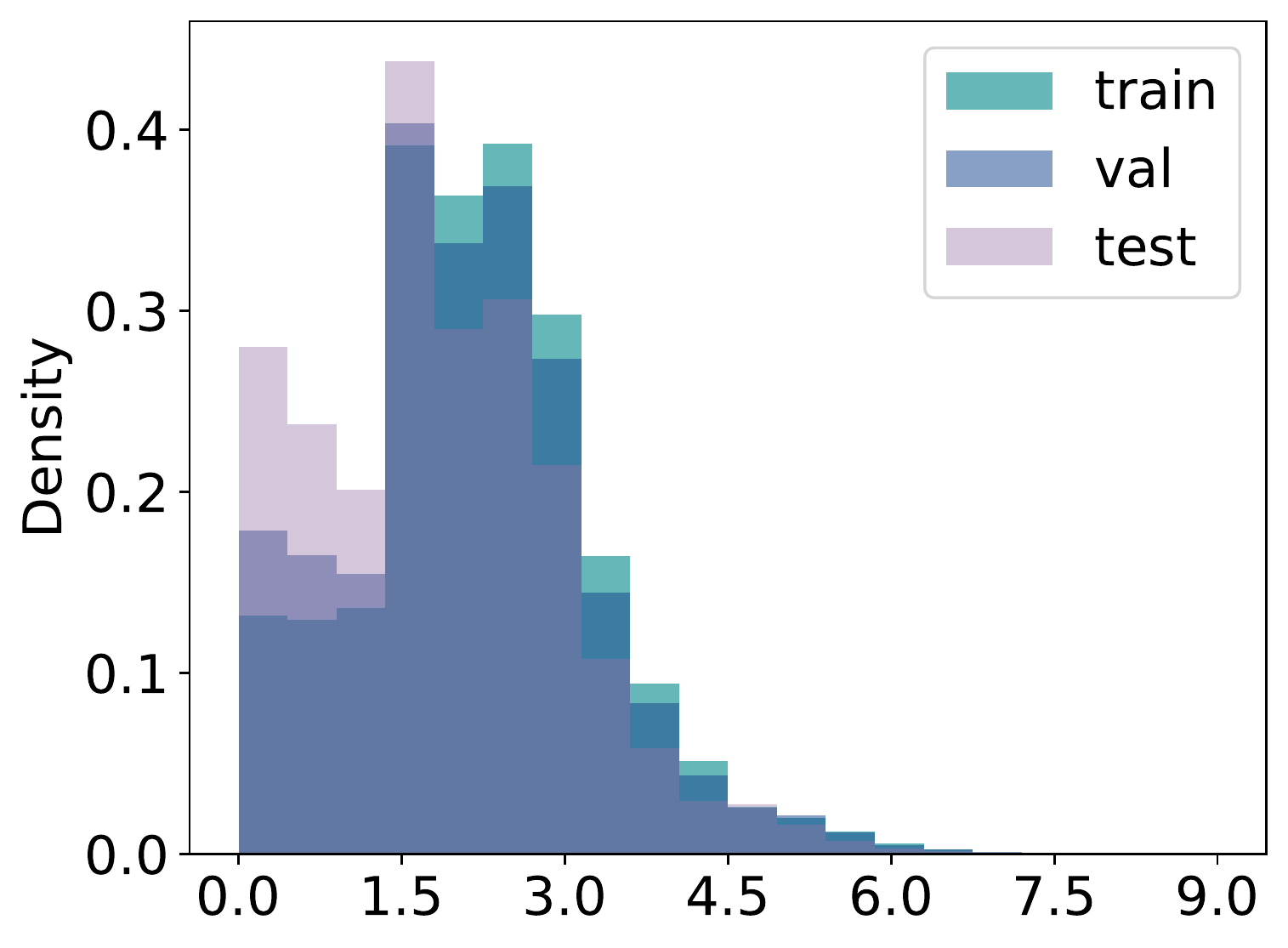}%
		\label{fig:dblp}}
  \caption{Distributions of potential citation counts (log).}
\label{fig:data}
\end{figure*}

The details of the two datasets are shown in Table \ref{tab:sample_count}. As aforementioned, from Figure \ref{fig:data} we can discover that after log transformation, the distributions of potential citation counts for all sets in both datasets are close to normal distribution. 

\subsection{Baseline models}
The baseline models are mainly divided into specialized citation prediction models, cascade prediction models, and dynamic GNN models.

\textbf{Citation prediction models:}
We apply recent citation prediction models as baselines.

\begin{itemize}
    \item \textbf{SciBERT} \citep{beltagy2019scibert} directly uses the SciBERT to encode the target paper's title and abstract and to estimate potential citation count/interval based on the representation of [CLS] token.
    \item \textbf{DGCBERT} \citep{xue2023dgcbert} is a paper acceptance prediction model built on SciBERT. The model constructs two distinct graphs based on the self-attention matrices, and then utilizes APPNP-based graph convolution to capture the lexical and semantic-level features of the paper content. Besides, it can mitigate the over-smoothing problem of SciBERT. In this work, we complement the classification MLP with a regression MLP.
    \item \textbf{HINTS} \citep{jiang2021hints} encodes heterogeneous graphs with R-GCN to capture the different relationships among papers, authors, venues, and keywords. To address the cold-start problem with no accumulated citations, it focuses on extracting features of metadata, including references. After utilizing GRU to aggregate graph information at each time step, it employs a decoder to forecast citation counts for each subsequent year. To fit the setting of our work, we take into account the year span between the observation year and the predicted year, and then sum the citation counts of all predicted years. Since not all datasets contain keywords, we remove the keyword-related information. For PMC, we also remove the author-related information due to the lack of author ids. More importantly, we replace the randomized ID embedding with text embedding the same as ours, which achieves better performance than the original HINTS.
\end{itemize}

\textbf{Cascade prediction models:}
To forecast the popularity (citation count) of papers based on citation cascade graphs, we utilize both traditional and contemporary cascade prediction models. We just use the most recent N years prior to the anticipated observation time rather than the time after publication to adapt these models to our emphasized real-world situation. 

\begin{itemize}
    \item \textbf{DeepHawkes} \citep{cao2017deephawkes} makes use of Hawkes point processes and RNNs for cascade prediction. Unlike DeepCas \citep{li2017deepcas}, it considers the group of authors as one user in the cascade graphs and transforms the raw citation networks into citation cascades.
    \item \textbf{GTGCN} \citep{yang2022gtgcn} utilizes the temporal encoder incorporating temporal embedding to encode temporal information. It begins by extracting representations of each paper via BiGCN. Then, the time sequence is fed into GTGRU, which integrates a temporal encoder into GRU module to capture the time-decay effect. The final valid layer of BiGCN and GTGRU are concatenated to predict the final popularity.
    \item \textbf{CCGL} \citep{xu2022ccgl} provides a novel method to augment the cascade graphs for contrastive learning and highlights the importance of utilizing unlabeled data. Moreover, it fine-tunes the model on the labeled data and further distills the model to avoid model bias. However, we argue that using unlabeled data should not include cascade graphs beyond observation time to avoid data leakage. Instead, we take the paper that publishes several years (like 5 years) before the observation time point as unlabeled data.
    \item \textbf{MUCas} \citep{chen2022mucas} first divides the cascade graphs into snapshots with time interval-aware sampling. Then, it makes use of MUG-caps to extract representations through the capsule networks at the order-, node- and graph-level. Finally, it aggregates the representations of different time periods by using a sub-graph level influence attention.
\end{itemize}

\textbf{Dynamic GNN models:}
We apply recent dynamic GNN models to extract paper representations and take the impact prediction as a node regression and classification task.

\begin{itemize}
    \item \textbf{Dysat} \citep{sankar2020dysat} utilizes multiple self-attention modules to encode structural and temporal information. It first uses structural self-attention to model structural information across different years and then re-weights temporal information across years using temporal self-attention.
    \item \textbf{EvolveGCN} \citep{pareja2020evolvegcn} integrates RNN structure into GNN modules. It considers graph structures as hidden states of GRU and takes the parameters of GRU as GCN parameters. It first obtains the graph embedding by multiplying the node feature matrix with the adjacent matrix, and then inputs the graph embeddings year by year like RNN to get the final embedding of the graph.
    \item \textbf{ROLAND} \citep{you2022roland} is a novel dynamic GCN model that inherits static GCN methods and integrates edge features. Unlike dynamic GCN methods that just take the final output of multiple GCN layers as input of RNN, ROLAND accepts all the output of multiple GCN layers as hierarchical input of RNN, to better learn temporal information.
    \item \textbf{H$^2$GCN} is the base model of our proposed model. It utilizes GCN for graph representation learning and then uses a sequential model transformer to encode the temporal graph sequence.
\end{itemize}

\subsection{Implementation details}
We implement all baselines and our proposed model with PyTorch. These models are trained on one NVIDIA A100 40GB GPU. All the baselines use recommended settings and are trained with Adam optimizer \citep{kingma2014adam}. We evaluate the models with the validation set and choose the model with the best MALE as the final model. 

We use DGL to construct our model. The settings of hyper-parameters inside our model are the same for both datasets. We set the number of GNN layers $L$ to 4, the interval prediction weight $\alpha$ to 0.5, the contrastive learning weight $\beta$ to 0.5, the hard negative samples $n^h$ to 2, and the augmentation rate $p^d$ to 0.1. To train our proposed model, we set the learning rate to 1e-4, batch size to 32, and epochs to 30 for PMC and 20 for DBLP.

For potential citation count prediction, we use MALE and LogR$^2$ as evaluation metrics. Due to the imbalance of labels in the interval classification task, we use Macro-F1 to complement the accuracy.

\section{Results and analysis}
\label{sec:results}

\begin{table}[htbp]
  \centering
  \caption{Experimental results of performance comparison of H$^2$CGL with alternative models in potential citation count on PMC and DBLP. We divide the results into three categories: previously-published papers, freshly-published papers, and total papers. The best results are shown in bold, and the second results are underlined. Significant improvements over best baseline results are marked with $^*$.}
    \begin{tabular}{c|l|rr|rr|rr}
    \toprule
    \multirow{2}[4]{*}{\textbf{PMC}} & \multicolumn{1}{c|}{\multirow{2}[4]{*}{model}} & \multicolumn{2}{c|}{previous} & \multicolumn{2}{c|}{fresh} & \multicolumn{2}{c}{total} \\
\cmidrule{3-8}          &       & \multicolumn{1}{c}{MALE$\downarrow$} & \multicolumn{1}{c|}{LogR$^2\uparrow$} & \multicolumn{1}{c}{MALE$\downarrow$} & \multicolumn{1}{c|}{LogR$^2\uparrow$} & \multicolumn{1}{c}{MALE$\downarrow$} & \multicolumn{1}{c}{LogR$^2\uparrow$} \\
    \midrule
    \multirow{3}[2]{*}{citation} & SciBERT & 0.5063  & 0.4273  & 0.5278  & -0.0943  & 0.5226  & 0.1266  \\
          & DGCBERT & 0.4807  & 0.4855  & 0.5229  & -0.0667  & 0.5126  & 0.1640  \\
          & HINTS & 0.7355  & -0.1638  & 0.5364  & -0.1853  & 0.5852  & -0.1383  \\
    \midrule
    \multirow{5}[2]{*}{cascade} & DeepHawkes & 0.5454  & 0.2954  & 0.4662  & 0.0558  & 0.4856  & 0.1719  \\
          & GTGCN & 0.6514  & 0.0618  & 0.5246  & -0.2537  & 0.5557  & -0.1004  \\
          & CCGL & 0.5503  & 0.3247  & 0.5256  & -0.1573  & 0.5316  & 0.0517  \\
          & MUCas & \textbf{0.4624} & \underline{0.5200}  & \underline{0.4284}  & \underline{0.2351}  & \underline{0.4367}  & \underline{0.3613}  \\
    \midrule
    \multirow{3}[2]{*}{DGNN} & Dysat & 0.6667  & 0.0397  & 0.4913  & 0.0223  & 0.5342  & 0.0610  \\
          & EGCN & 0.5759  & 0.2797  & 0.5224  & -0.1078  & 0.5355  & 0.0661  \\
          & ROLAND & 0.8154  & -0.2531  & 0.5085  & -0.0208  & 0.5837  & -0.0690  \\
    \midrule
    \multirow{3}[4]{*}{ours} & H$^2$GCN & 0.4850  & 0.4839  & 0.4394  & 0.2152  & 0.4505  & 0.3363  \\
          & H$^2$CGL & \underline{0.4627}  & \textbf{0.5207$^*$} & \textbf{0.4114$^*$} & \textbf{0.3222$^*$} & \textbf{0.4239$^*$} & \textbf{0.4150$^*$} \\
\cmidrule{2-8}          & \#Improve & \multicolumn{1}{c}{--} & 0.15\%$\uparrow$ & 3.98\%$\downarrow$ & 37.04\%$\uparrow$ & 2.94\%$\downarrow$ & 14.85\%$\uparrow$ \\
    \midrule
    \multirow{2}[4]{*}{\textbf{DBLP}} & \multicolumn{1}{c|}{\multirow{2}[4]{*}{model}} & \multicolumn{2}{c|}{previous} & \multicolumn{2}{c|}{fresh} & \multicolumn{2}{c}{total} \\
\cmidrule{3-8}          &       & \multicolumn{1}{c}{MALE$\downarrow$} & \multicolumn{1}{c|}{LogR$^2\uparrow$} & \multicolumn{1}{c}{MALE$\downarrow$} & \multicolumn{1}{c|}{LogR$^2\uparrow$} & \multicolumn{1}{c}{MALE$\downarrow$} & \multicolumn{1}{c}{LogR$^2\uparrow$} \\
    \midrule
    \multirow{3}[2]{*}{citation} & SciBERT & 0.8662  & 0.1639  & 0.6986  & -0.1686  & 0.8267  & 0.2610  \\
          & DGCBERT & 0.8532  & 0.1682  & 0.7247  & -0.2931  & 0.8229  & 0.2496  \\
          & HINTS & 0.8036  & 0.2020  & 0.6376  & -0.0793  & 0.7645  & 0.2990  \\
    \midrule
    \multirow{5}[2]{*}{cascade} & DeepHawkes & 0.7285  & 0.3556  & 0.5573  & 0.1572  & 0.6881  & 0.4372  \\
          & GTGCN & 1.0229  & -0.3070  & 0.6257  & -0.1440  & 0.9292  & -0.0752  \\
          & CCGL & 0.8114  & 0.2043  & 0.6836  & -0.2738  & 0.7812  & 0.2778  \\
          & MUCas & \underline{0.5273}  & \underline{0.6465}  & \underline{0.4581}  & \underline{0.4406}  & \underline{0.5110}  & \underline{0.6799}  \\
    \midrule
    \multirow{2}[2]{*}{dynamic} & EGCN & 0.9192  & -0.0180  & 0.7350  & -0.2701  & 0.8758  & 0.1182  \\
          & ROLAND & 1.3471  & -0.8481  & 0.9322  & -1.1761  & 1.2492  & -0.5857  \\
    \midrule
    \multirow{3}[4]{*}{ours} & H$^2$GCN & 0.8534  & 0.1537  & 0.8439  & -0.5610  & 0.8511  & 0.2077  \\
          & H$^2$CGL & \textbf{0.4623$^*$} & \textbf{0.7355$^*$} & \textbf{0.4274$^*$} & \textbf{0.5325$^*$} & \textbf{0.4540$^*$} & \textbf{0.7548$^*$} \\
\cmidrule{2-8}          & \#Improve & 12.33\%$\downarrow$ & 13.76\%$\uparrow$ & 6.70\%$\downarrow$ & 20.86\%$\uparrow$ & 11.14\%$\downarrow$ & 11.00\%$\uparrow$ \\
    \bottomrule
    \end{tabular}

  \label{tab:reg}
\end{table}

\begin{table}[htbp]
  \centering
  \caption{Experimental results of performance comparison of H$^2$CGL with alternative models in potential interval prediction (\%) on PMC and DBLP. We divide the results into three categories: previously-published papers, freshly-published papers, and total papers. The best results are shown in bold, and the second results are underlined. Significant improvements over best baseline results are marked with $^*$.}
    \begin{tabular}{c|l|rr|rr|rr}
    \toprule
    \multirow{2}[4]{*}{\textbf{PMC}} & \multicolumn{1}{c|}{\multirow{2}[4]{*}{model}} & \multicolumn{2}{c|}{previous} & \multicolumn{2}{c|}{fresh} & \multicolumn{2}{c}{total} \\
\cmidrule{3-8}          &       & \multicolumn{1}{c}{Acc$\uparrow$} & \multicolumn{1}{c|}{F1$\uparrow$} & \multicolumn{1}{c}{Acc$\uparrow$} & \multicolumn{1}{c|}{F1$\uparrow$} & \multicolumn{1}{c}{Acc$\uparrow$} & \multicolumn{1}{c}{F1$\uparrow$} \\
    \midrule
    \multirow{3}[2]{*}{citation} & SciBERT & 74.63 & 69.07 & \underline{59.79} & 45.69 & \underline{63.43} & 52.83 \\
          & DGCBERT & \textbf{76.55} & \underline{70.31} & 59.16 & 44.11 & 63.42 & 52.12 \\
          & HINTS & 53.50 & 35.86 & 52.61 & 34.59 & 52.83 & 35.19 \\
    \midrule
    \multirow{5}[2]{*}{cascade} & DeepHawkes & 53.40 & 23.21 & 34.52 & 17.11 & 39.14 & 18.75 \\
          & GTGCN & 53.40 & 23.21 & 34.52 & 17.11 & 39.14 & 18.75 \\
          & CCGL & 60.80 & 40.10 & 44.83 & 28.98 & 48.74 & 32.49 \\
          & MUCas & 73.82 & 58.26 & 53.06 & 50.28 & 58.14 & 52.38 \\
    \midrule
    \multirow{3}[2]{*}{DGNN} & Dysat & 58.10 & 37.95 & 53.74 & 35.40 & 54.81 & 36.71 \\
          & EGCN & 63.30 & 49.51 & 51.77 & 35.98 & 54.60 & 39.52 \\
          & ROLAND & 48.32 & 39.68 & 56.22 & 38.76 & 54.29 & 40.77 \\
    \midrule
    \multirow{3}[4]{*}{ours} & H$^2$GCN & 74.23 & 67.56 & 55.02 & \underline{52.49} & 59.73 & \underline{56.50} \\
          & H$^2$CGL & \underline{74.81} & \textbf{70.51$^*$} & \textbf{62.88$^*$} & \textbf{57.53$^*$} & \textbf{65.80$^*$} & \textbf{62.00$^*$} \\
\cmidrule{2-8}          & \#Improve & \multicolumn{1}{c}{--} & 0.29\%$\uparrow$ & 5.16\%$\uparrow$ & 9.59\%$\uparrow$ & 3.74\%$\uparrow$ & 9.74\%$\uparrow$ \\
    \midrule
    \multirow{2}[4]{*}{\textbf{DBLP}} & \multicolumn{1}{c|}{\multirow{2}[4]{*}{model}} & \multicolumn{2}{c|}{previous} & \multicolumn{2}{c|}{fresh} & \multicolumn{2}{c}{total} \\
\cmidrule{3-8}          &       & \multicolumn{1}{c}{Acc$\uparrow$} & \multicolumn{1}{c|}{F1$\uparrow$} & \multicolumn{1}{c}{Acc$\uparrow$} & \multicolumn{1}{c|}{F1$\uparrow$} & \multicolumn{1}{c}{Acc$\uparrow$} & \multicolumn{1}{c}{F1$\uparrow$} \\
    \midrule
    \multirow{3}[2]{*}{citation} & SciBERT & 73.50 & 68.56 & 57.53 & 44.44 & 69.74 & 64.45 \\
          & DGCBERT & 71.05 & 63.87 & 56.93 & 44.99 & 67.72 & 60.99 \\
          & HINTS & 67.42 & 45.01 & 52.09 & 36.52 & 63.81 & 43.55 \\
    \midrule
    \multirow{5}[2]{*}{cascade} & DeepHawkes & 84.40 & 46.72 & 55.99 & 37.79 & 77.70 & 45.63 \\
          & GTGCN & 75.48 & 39.65 & 43.16 & 20.53 & 67.86 & 34.44 \\
          & CCGL & 81.99 & 59.00 & 46.32 & 31.11 & 73.58 & 51.13 \\
          & MUCas & \underline{87.58} & \underline{71.17} & \underline{68.56} & \underline{66.44} & \underline{83.09} & \underline{72.08} \\
    \midrule
    \multirow{2}[2]{*}{dynamic} & EGCN & 84.90 & 46.40 & 45.64 & 24.37 & 75.64 & 39.72 \\
          & ROLAND & 47.51 & 38.65 & 43.53 & 22.32 & 46.57 & 36.30 \\
    \midrule
    \multirow{3}[4]{*}{ours} & H$^2$GCN & 70.59 & 60.44 & 53.45 & 23.22 & 66.55 & 57.64 \\
          & H$^2$CGL & \textbf{89.30$^*$} & \textbf{83.41$^*$} & \textbf{69.02$^*$} & \textbf{67.51$^*$} & \textbf{84.52$^*$} & \textbf{80.50$^*$} \\
\cmidrule{2-8}          & \#Improve & 1.96\%$\uparrow$ & 17.19\%$\uparrow$ & 0.67\%$\uparrow$ & 1.61\%$\uparrow$ & 1.71\%$\uparrow$ & 11.68\%$\uparrow$ \\
    \bottomrule
    \end{tabular}%
  \label{tab:cls}%
\end{table}%

\subsection{Performance Comparison (RQ1)}

Table \ref{tab:reg} and \ref{tab:cls} illustrate the performance of H$^2$CGL and all baselines. From the results, we can observe:

(1) Indeed, the discrepancy between the PMC and DBLP datasets arises from the distinctive characteristics of their respective fields of study and variations in data quality. The PMC dataset may pose greater challenges than DBLP due to the absence of detailed author information and the significant scale gap between the training and test sets, particularly in the citation interval task. Since several models can achieve comparable performance in both datasets, the task remains challenging. For example, in citation count prediction, the ranking of citation models (LogR$^2$) in DBLP is DGCBERT $>$ SciBERT $>$ HINTS, but the ranking changes oppositely in PMC: HINTS $>$ SciBERT $>$ DGCBERT. This suggests that without access to author information, HINTS may not be able to extract reliable representations for accurate predictions. Our proposed model exhibits excellent performance for both PMC and DBLP datasets, showcasing its generality. Remarkably, even when lacking author information in PMC, the model remains robust and outperforms all baseline methods in almost all metrics.

(2) In general, most models perform better on predictions for previous papers than fresh papers. However, they can achieve slightly better performance on fresh papers in terms of MALE, just because their absolute potential citations are lower than previous ones. Our proposed model has shown superior performance compared to other models on most metrics, particularly for fresh papers that are not presented during the training phase, achieving a more than 20\% improvement in LogR$^2$ for both datasets. This result indicates that our model can effectively learn useful patterns that can be generalized to new papers, which demonstrates its potential for real-world applications.

(3) Content-based models like SciBERT and DGCBERT can encode semantic information from the title and abstract of papers, leading to decent performances in citation interval classification. Especially, in PMC where training samples are limited, DGCBERT achieves the best Acc (76.55\%) for previous papers, while SciBERT ranks second for both fresh papers (59.79\%) and total (63.43\%). In contrast, cascade prediction models are able to model the information diffusion process. For citation count prediction, although DeepHawkes is a very simple cascade prediction model, it still ranks third in PMC (0.4856), and second in DBLP (0.6881) among all baselines. Thus, content-based and cascade prediction models focus on different aspects of the impact prediction task. MUCas may be the best baseline for both datasets. It only encodes the structural and temporal information, so it can be easily transferred to new papers with similar patterns. However, due to the lack of semantic information, it performs terribly in the interval prediction for the PMC dataset. 

(4) General dynamic GNNs like Dysat (Out of memory on DBLP), EGCN, and ROLAND encode entire citation networks. They perform worse than other baselines, especially on DBLP. Taking EGCN as an example, in both datasets, its LogR$^2$ is close to 0.1 for count prediction, while its F1 is about 39.5\% for interval prediction. There exists an exponential gap between them and other remarkable models (more than 0.3 in LogR$^2$ and 60\% in F1). As entire citation networks always contain noise information irrelevant to the target papers, they cannot model predicted samples attentively. Additionally, the huge size of the complete citation network may always be a big challenge for them. In reality, the citation network may be much larger, and these models cannot tackle this challenge.

(5) Heterogeneous information is crucial in the citation network. HINTS and H$^2$GCN simulate the heterogeneous citation network and have shown promising performance. Although ROLAND also encodes heterogeneous information, it models them in a homogeneous manner with the edge embedding, losing important heterogeneous features. H$^2$GCN further performs better than HINTS, which demonstrates the significance of target-centered information. However, it performs well on PMC (comparable to MUCas) while facing a huge drop in DBLP datasets (cannot defeat SciBERT in both citation count and interval prediction). The graph per paper in DBLP is more complex than that in PMC. Without the mechanisms of adaptive aggregation like virtual snapshot node and R-GAT, the H$^2$GCN still suffers from the noise of irrelevant information.

(6) Our proposed model H$^2$CGL outperforms all the baselines and achieves the best results in almost all metrics. Specifically, except for the previous papers in PMC, H$^2$CGL surpasses all the baselines by a large margin. In the case of fresh papers, the H$^2$CGL improves by 3.98\%, 6.70\% in MALE and 37.04\%, 20.86\% in LogR$^2$ for citation count prediction, while 5.16\%, 0.67\% in Acc and 9.59\%, 1.61\% in F1 for citation interval prediction, comparing with the best baseline MUCas. These results demonstrate that our model can effectively harness newly-coming information, which is more practical in dynamic realistic situations. For previous papers with sufficient information in DBLP, our model can also boost 12.33\% in MALE and 13.76\% in LogR$^2$ for citation interval prediction, and 1.96\% in Acc and 17.19\% in F1. Our model can further address the issue of data imbalance across different classes, resulting in significant improvements in F1 for the citation interval prediction task. Compared with the content-based citation prediction models, it also extracts rich dynamic information from the citation network of the target papers by constructing hierarchical and heterogeneous graphs. Compared with the context-based prediction models, it can make use of the semantic information extracted from the content of papers and dynamics added by newly published papers. Particularly, the adjusted R-GCN encoder including C-GIN, R-GAT, and weighted GIN is advisable to exploit the semantic, structural, and temporal features simultaneously. Moreover, with the help of the contrastive learning module, H$^2$CGL can better distinguish papers with similar topics. 

\subsection{Ablation Test (RQ2)}

\begin{figure}[]
  \centering
  	\centering
	\subfloat[PMC]{\includegraphics[width=0.22\textwidth]{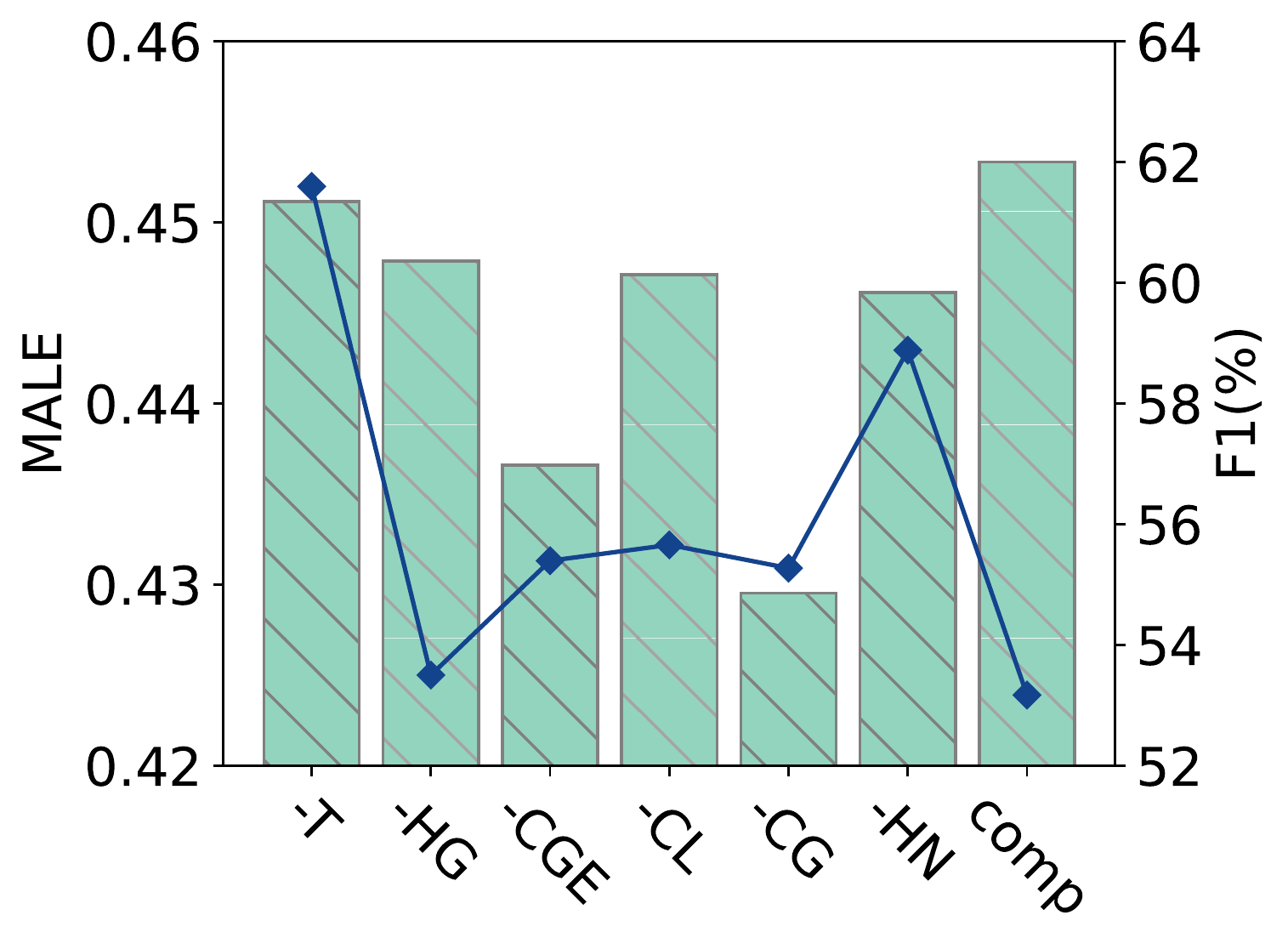}
		\label{fig:ab_pmc}}
        \hfil
	\subfloat[DBLP]{\includegraphics[width=0.22\textwidth]{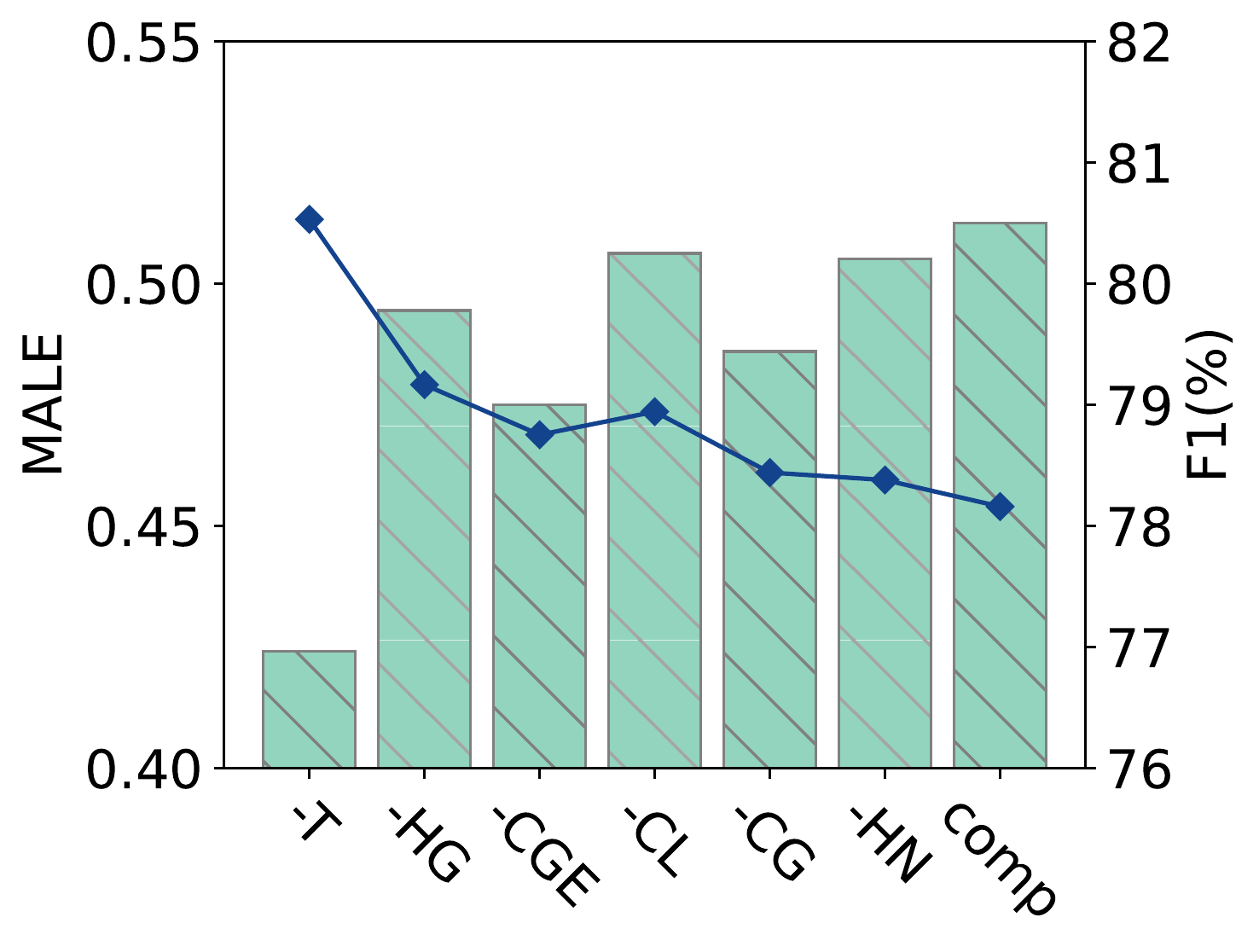}
		\label{fig:ab_dblp}}
  \caption{The performance of H$^2$CGL and its variants. Lines denote MALE and bars denote MA-F1.} 
\label{fig:ablation}
\end{figure}

We then conduct ablation tests of H$^2$CGL to validate the effectiveness of its modules. As shown in Figure \ref{fig:ablation}, (-T) replaces the text embeddings with the randomly initialized embeddings. (-HG) removes the hierarchical graph information, replacing the Weighted-GIN with a sequential model layer (self-attention layer). (-CGE) removes the citation-aware graph encoder that contains C-GIN and R-GAT, replacing them with the original version of GIN and GAT. (-CL) removes the contrastive learning module. (-CG) replaces our CL module with common graph augmentation methods like edge drop and attribute mask. (-HN) utilizes graph representations of other samples in batches instead of the selected hard negative samples. 

We can observe that in both datasets, the performance of H$^2$CGL decreases by removing each module. The semantic information is essential to our model. Without the hierarchical graph, the model faces difficulty in modeling the dynamics between different snapshots, though the transformer can be seen as modeling a fully-connected weighted graph. The citation-aware graph encoder can model the dynamic heterogeneous information from the perspective of scientific diffusion. Each part of the contrastive learning module is significant, including itself, the augmentation methods, and the hard negative sampling methods. The model severely degrades with the normal graph augmentation methods, since it will destroy important relations in the citation network. Additionally, without the hard negative sampling method, the performance of the model can not be improved, even compared with the model without contrastive learning in PMC. Since the content and context of papers are originally distinct among different topics, contrasting the target paper with papers from different topics cannot help the model learn distinct representation. We should realize that though the impact of papers is relevant to the topics, it is futile to shuffle the papers with their topics.

\subsection{Hype-parameters Test (RQ3)}

\begin{figure}[]
  \centering
  	\centering
	\subfloat[$L$]{\includegraphics[width=0.22\textwidth]{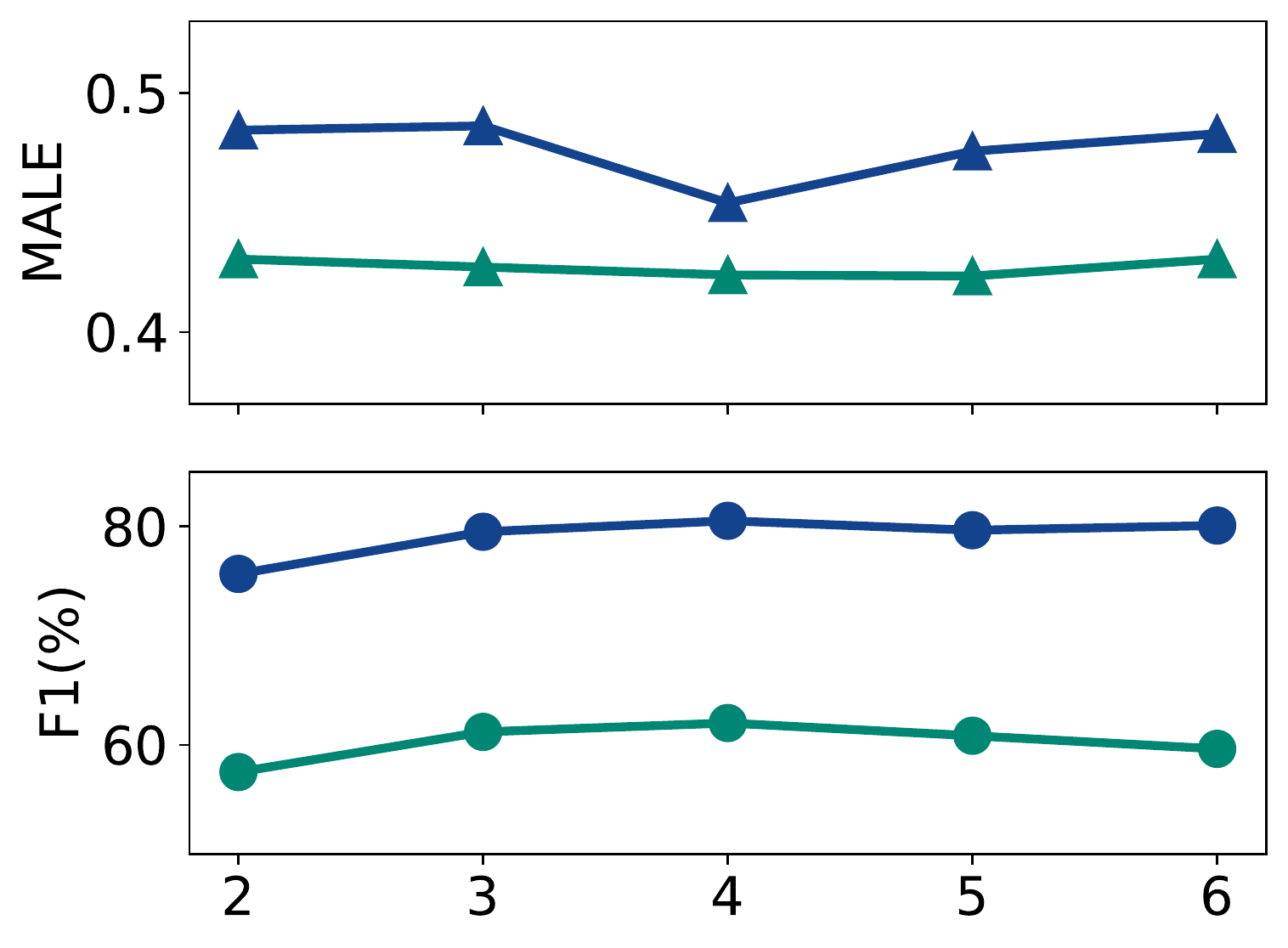}
		\label{fig:param_layer}}
        \hfil
	\subfloat[$n^t$]{\includegraphics[width=0.22\textwidth]{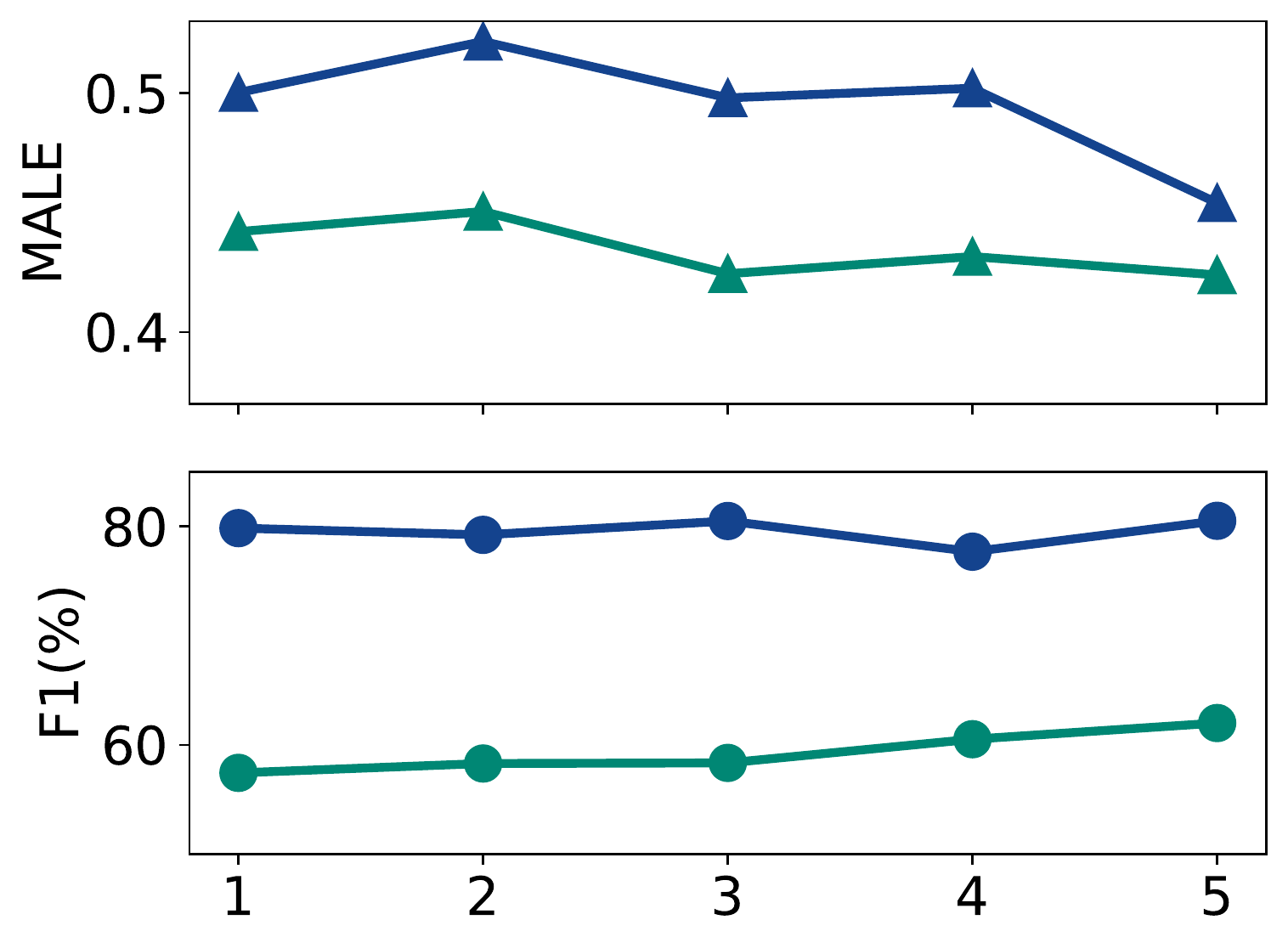}
		\label{fig:param_tw}}
          \hfil
	\subfloat[$\alpha$]{\includegraphics[width=0.22\textwidth]{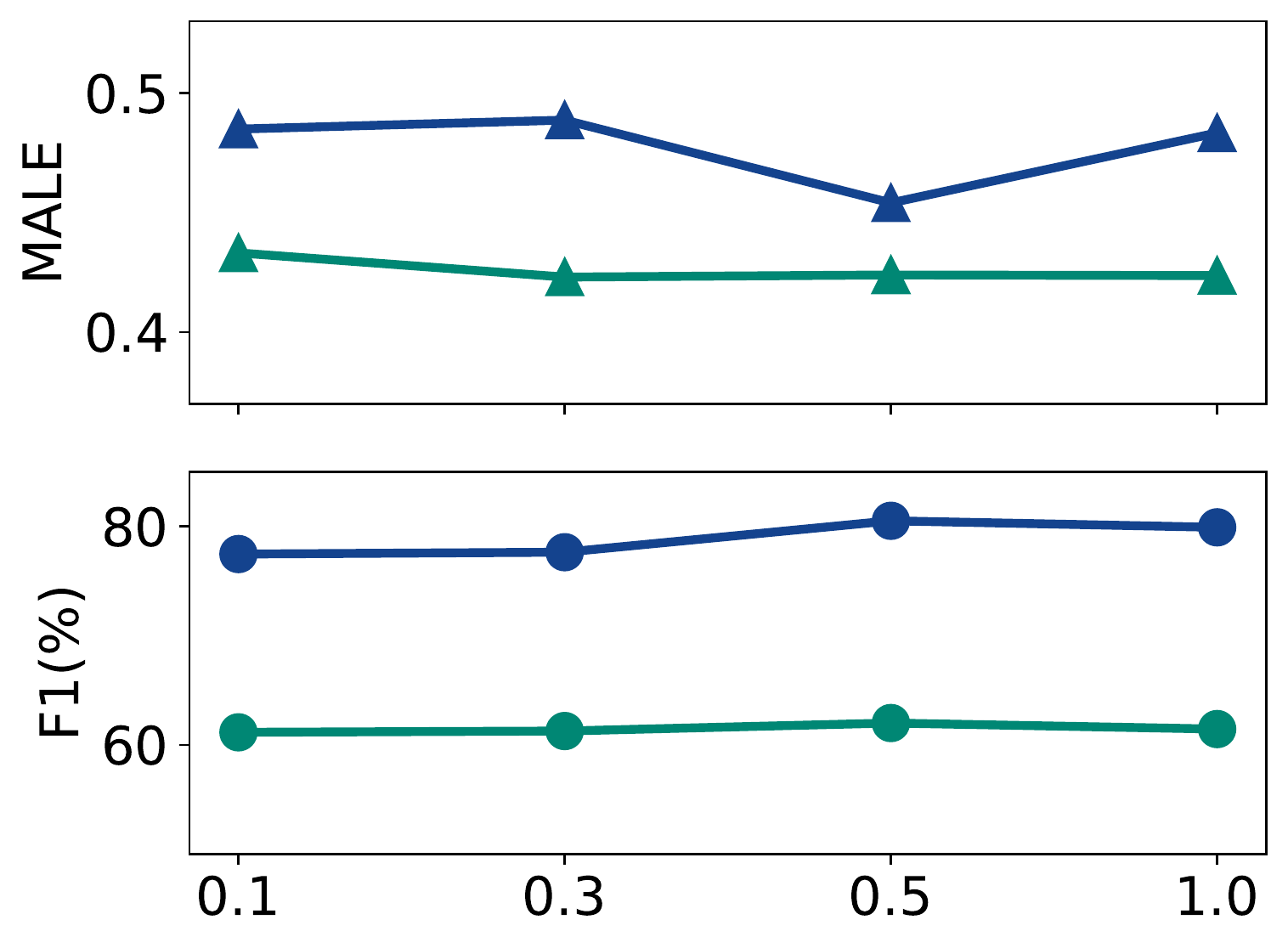}
		\label{fig:param_cls_w}}
          \hfil
	\subfloat[$\beta$]{\includegraphics[width=0.22\textwidth]{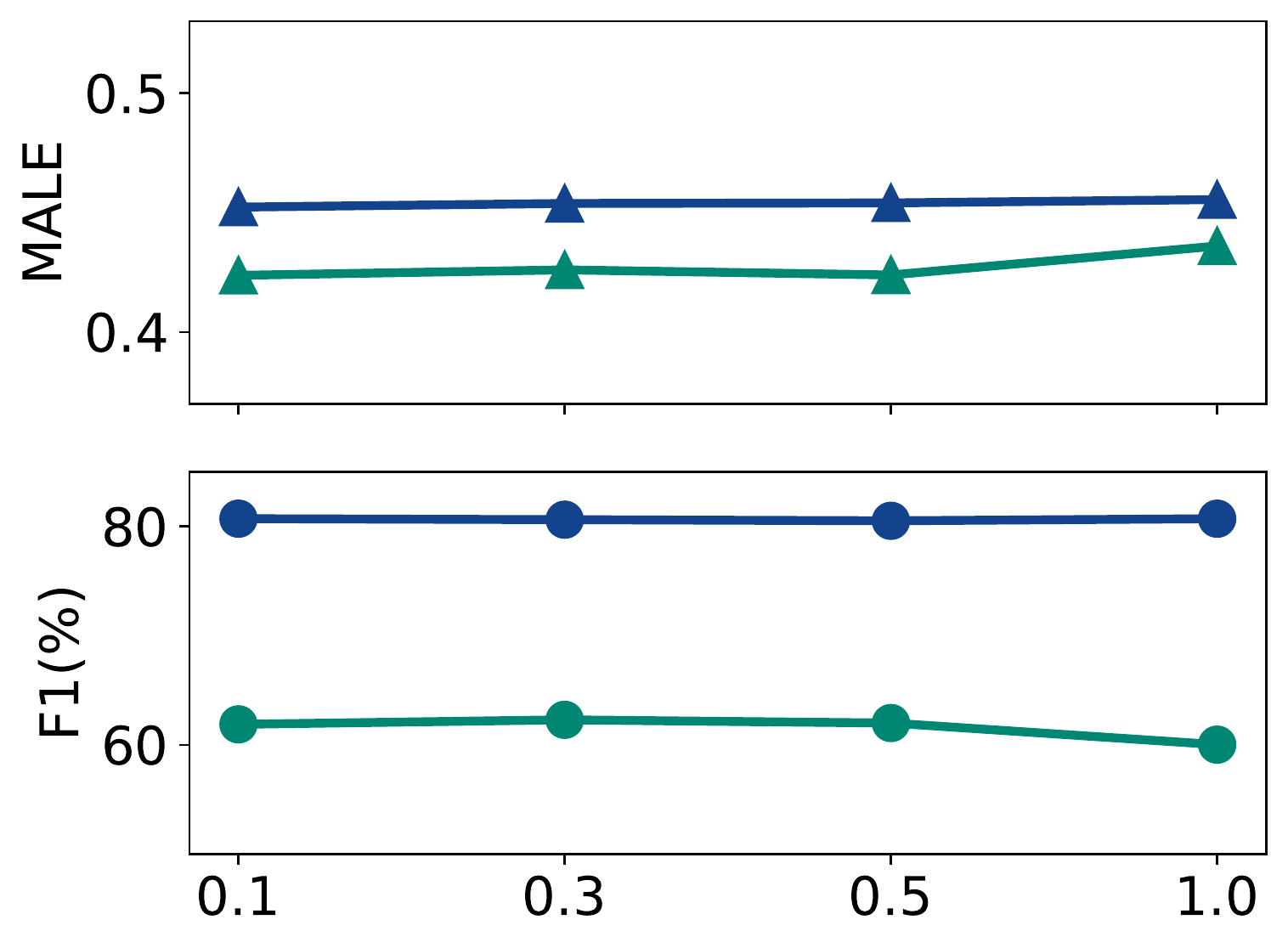}
		\label{fig:param_cl_w}}
  \caption{Hyper-parameters test of H$^2$CGL. The green lines denote PMC and blue lines denote DBLP. Triangles denote MALE and circles denote MA-F1.}
\label{fig:hyper}
\end{figure}

We also conduct hyper-parameter tests to show the robustness of H$^2$CGL. We consider the significance of the number of graph encoder layers $L$, the observation window length $n^t$, the interval prediction loss weight $\alpha$, and the contrast learning loss weight $\beta$.

Since we utilize GIN with the original information as the main prototype of GNN encoder, the model can stack many layers. For both of the datasets, the optimal number of layers is 4, which may be the balance point between structural and temporal information. The observation window length may be also significant since it decides how many years of dynamics are to be captured by our model. We can observe that the performance of our model correlates with the observation window length. Performance improves slightly when the window length is made longer. The weight of the citation interval prediction task is an important hyper-parameter. Both the performance of citation count and interval prediction tasks may benefit from the appropriate setting. Additionally, the weight of contrastive learning may not be very sensitive. The model demonstrates robustness with different CL weights. 

\subsection{Attention Visualization (RQ4)}
\label{sec:visual}
\begin{figure*}[]
  \centering
  	\centering
        \subfloat[C-GIN]{\includegraphics[width=0.22\textwidth]{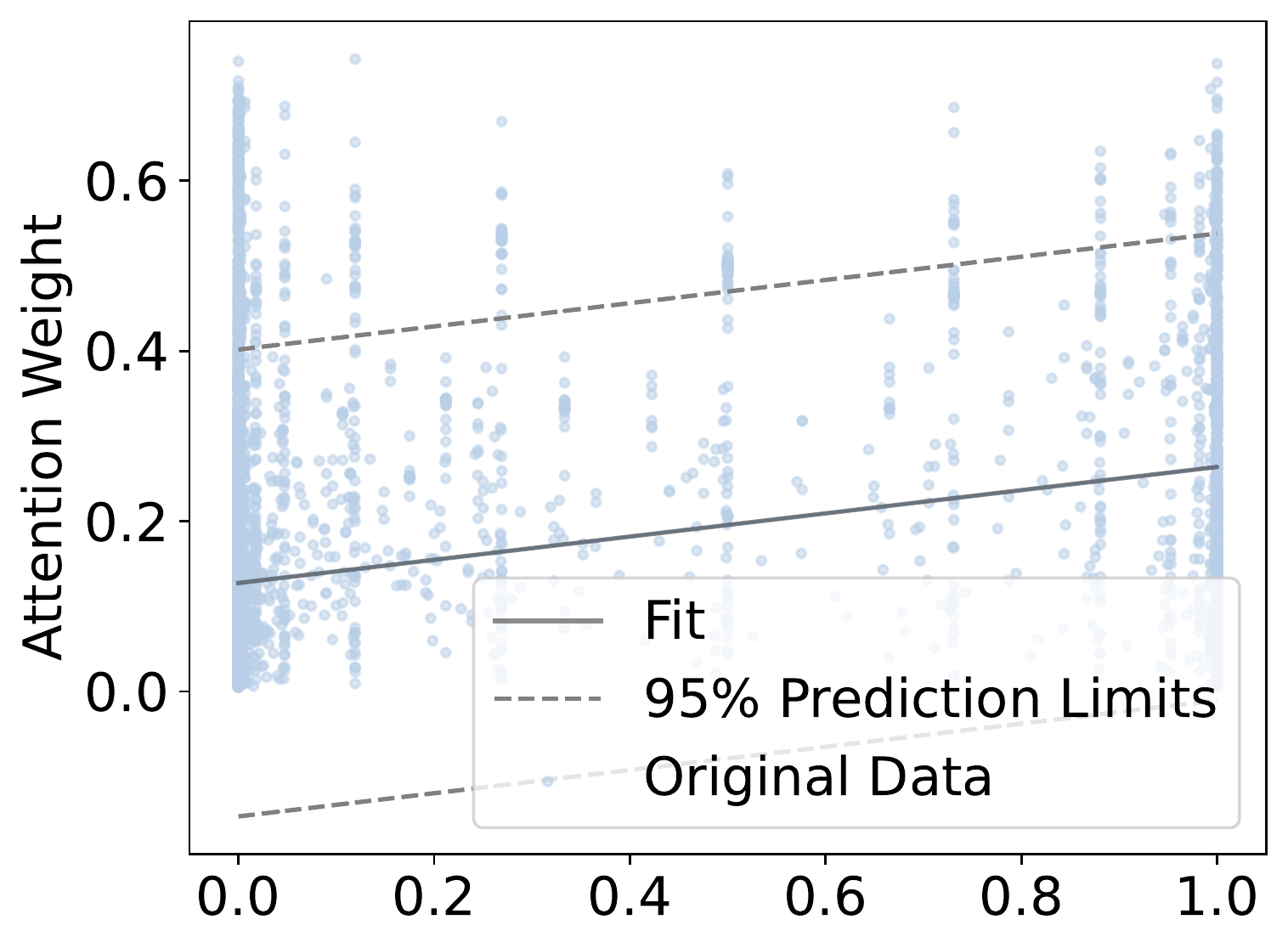}%
        \label{fig:c-gin}}
                \hfil
	\subfloat[R-GAT (2002)]{\includegraphics[width=0.22\textwidth]{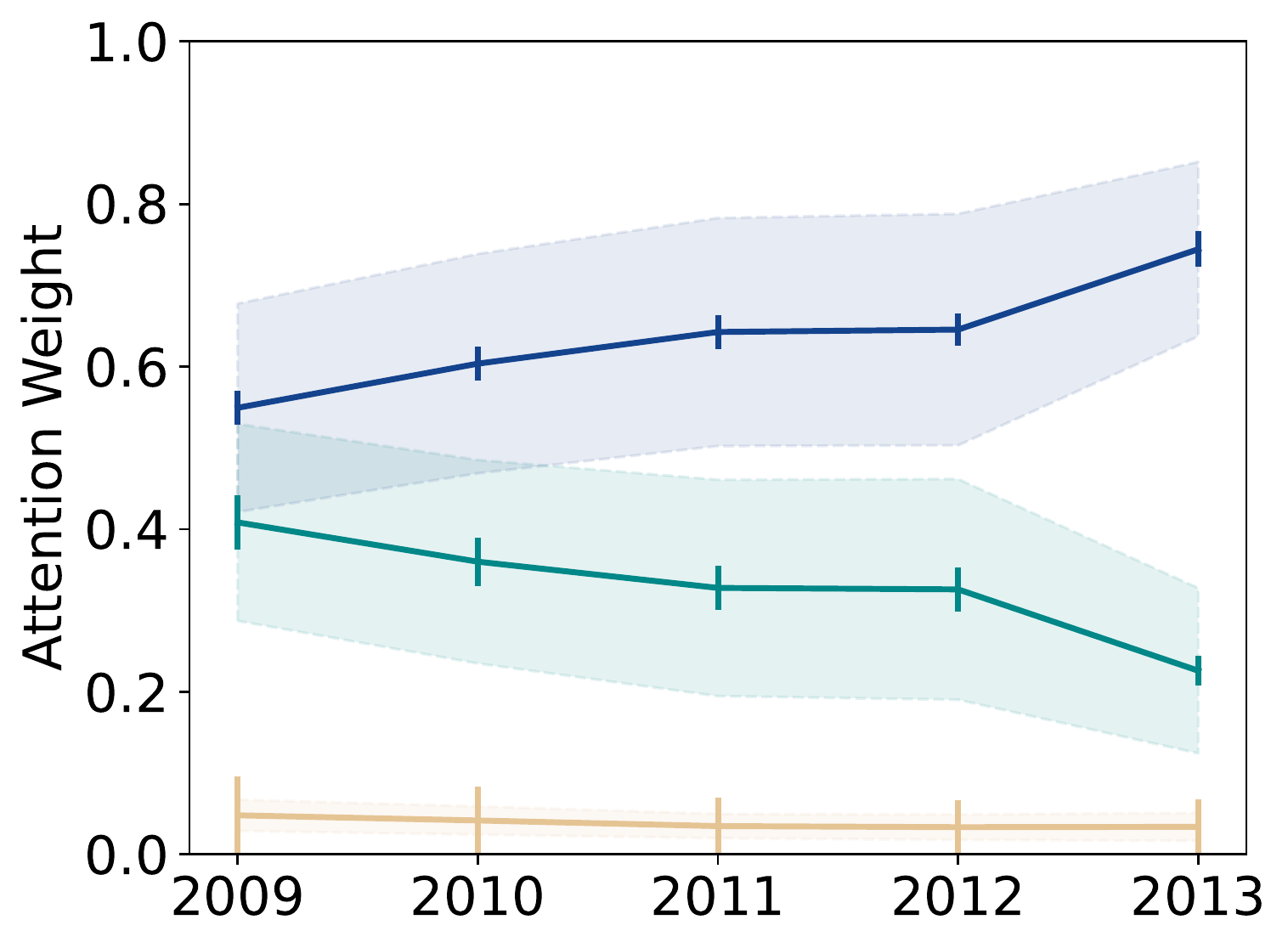}%
		\label{fig:previous}}
        \hfil
	\subfloat[R-GAT (2006)]{\includegraphics[width=0.22\textwidth]{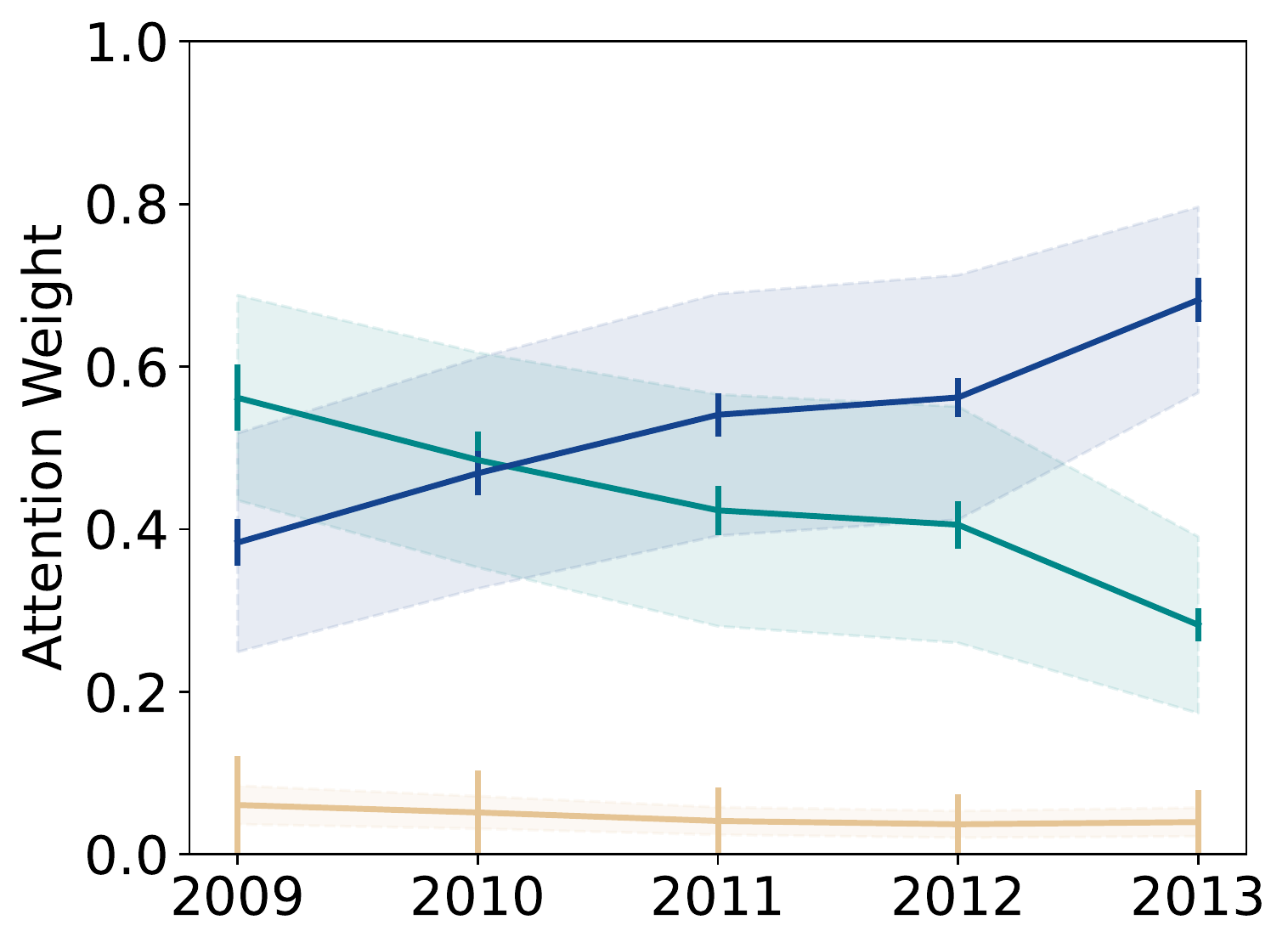}%
		\label{fig:common}}
        \hfil
  	\subfloat[R-GAT (2009)]{\includegraphics[width=0.22\textwidth]{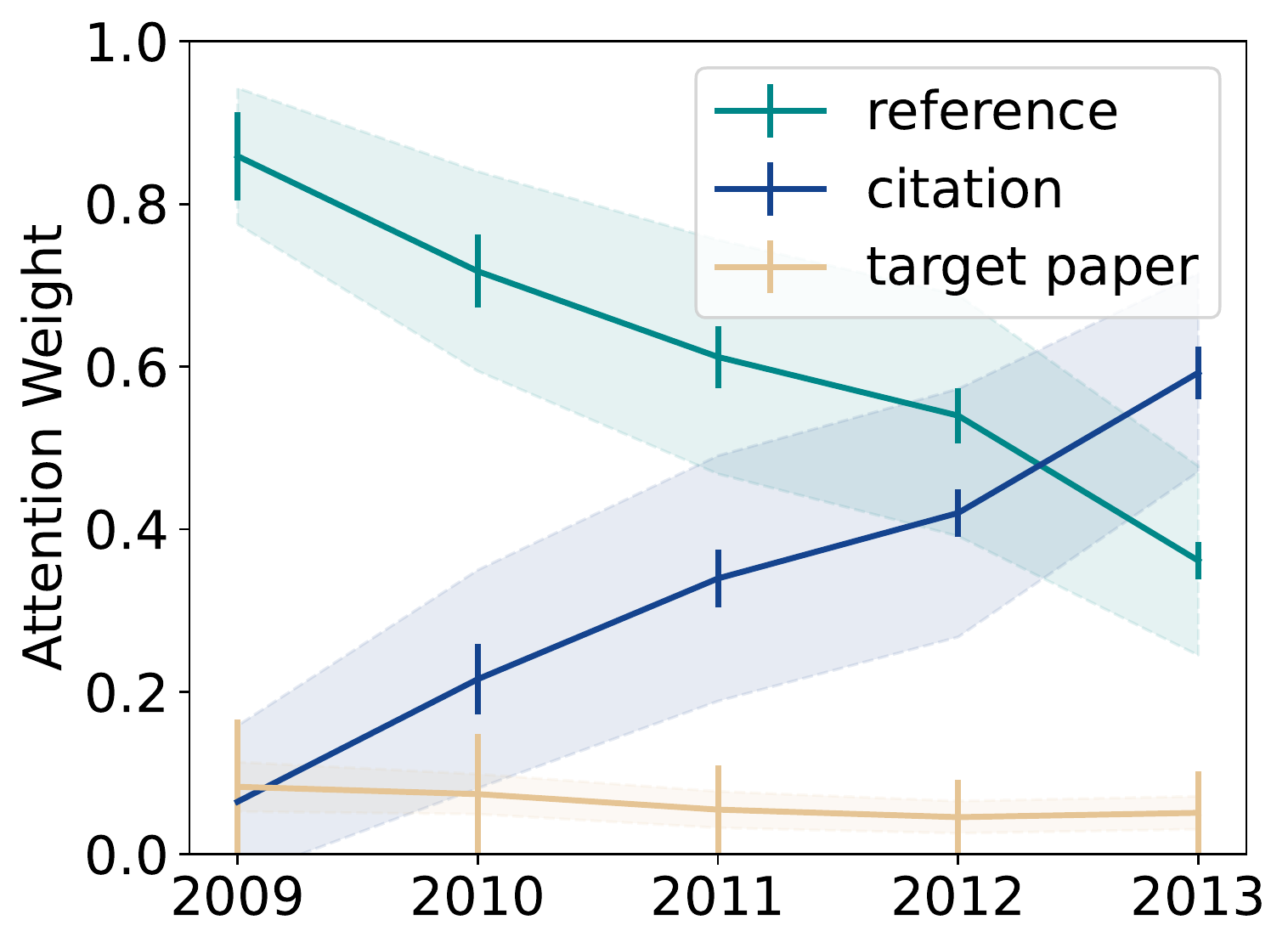}%
		\label{fig:fresh}}
  \caption{Visualization of attention weights in C-GIN and R-GAT on DBLP. In R-GAT, the transparent areas are standard deviations of different types, and the lengths of vertical lines represent attention weights average by paper counts.}
\label{fig:visualization}
\end{figure*}

We visualize the attention weights of C-GIN and R-GAT to provide insights into how these modules work on DBLP. 

For C-GIN, we first normalize citation counts of citing papers for each \textit{paper} node via softmax. Then, we fit a linear regression between the normalized numbers and their corresponding attentions in C-GIN. As illustrated in Figure \ref{fig:c-gin}, we observe a positive relationship between the attention weights and the citation counts. However, it is worth noting that the attention weights are not entirely dominated by the citation counts. The weights are distributed widely, indicating that C-GIN can also effectively capture the original information of papers.

For R-GAT, we performed a separate analysis on all papers published in 2002, 2006, and 2009 in the test set of DBLP. Figure \ref{fig:previous}, \ref{fig:common}, and \ref{fig:fresh} display the changes in attention weights associated with different paper types over the observation window. The trends observed in the attention weights of different source paper types illustrate how the interests of snapshot nodes evolve over time. For papers published in 2002, which represent previously-published papers, the importance of citation paper nodes outweighs that of reference paper nodes, as the target papers have accumulated enough citation papers to serve as a basis for evaluation. In addition, we can observe a turning point in 2012 where the types of snapshot nodes shift from common paper snapshots to previous paper snapshots. For previous paper snapshots, the importance of citation paper nodes is much greater compared to those in common paper snapshots. For papers published in 2006, they are common papers between previously-published and freshly-published papers. We can observe a cross point between citation paper nodes and reference paper nodes. Initially, the papers extract more information from the reference paper nodes. As time passes, the papers gradually have more citation paper nodes to encode, and eventually, citation paper nodes surpass the importance of reference paper nodes. The turning point in 2011 is also due to the change in the snapshot node type. Furthermore, the attention weights of the target paper nodes of common papers are slightly more important than those of previously-published papers. For papers published in 2009, which are representative of papers without accumulative citations, at first, they are dominated by reference paper nodes. Since they have few or even no citation paper nodes, the inherent information from references is much more important. With the passage of time, citation paper nodes become much more important not only in the total value but also in the average value. In 2013, both the total count and the average attention weight of citation paper nodes surpass those of reference paper nodes. Particularly, in 2009, although the count of citation paper nodes is larger than that of the target paper node, the target paper node is still more important than the citation paper nodes.

In conclusion, C-GIN and R-GAT can encode unique information within the dynamic citation network. For C-GIN, it will attach more attention to the highly-cited papers, but it will also determine the final importance from the perspective of the destination papers flexibly. For R-GAT, it will adaptively consider the significance of different types of source paper nodes based on the snapshot node types. For previously-published papers, citation paper nodes are the most important nodes. For common papers, there exists a turning point from the reference paper nodes to citation paper nodes. For papers without cumulative citations, the reference paper nodes dominate at first and are finally surpassed by the citation paper nodes. 

\subsection{Case Study (RQ4)}
\begin{figure*}[!h]
  \centering
  \includegraphics[width=0.9\textwidth]{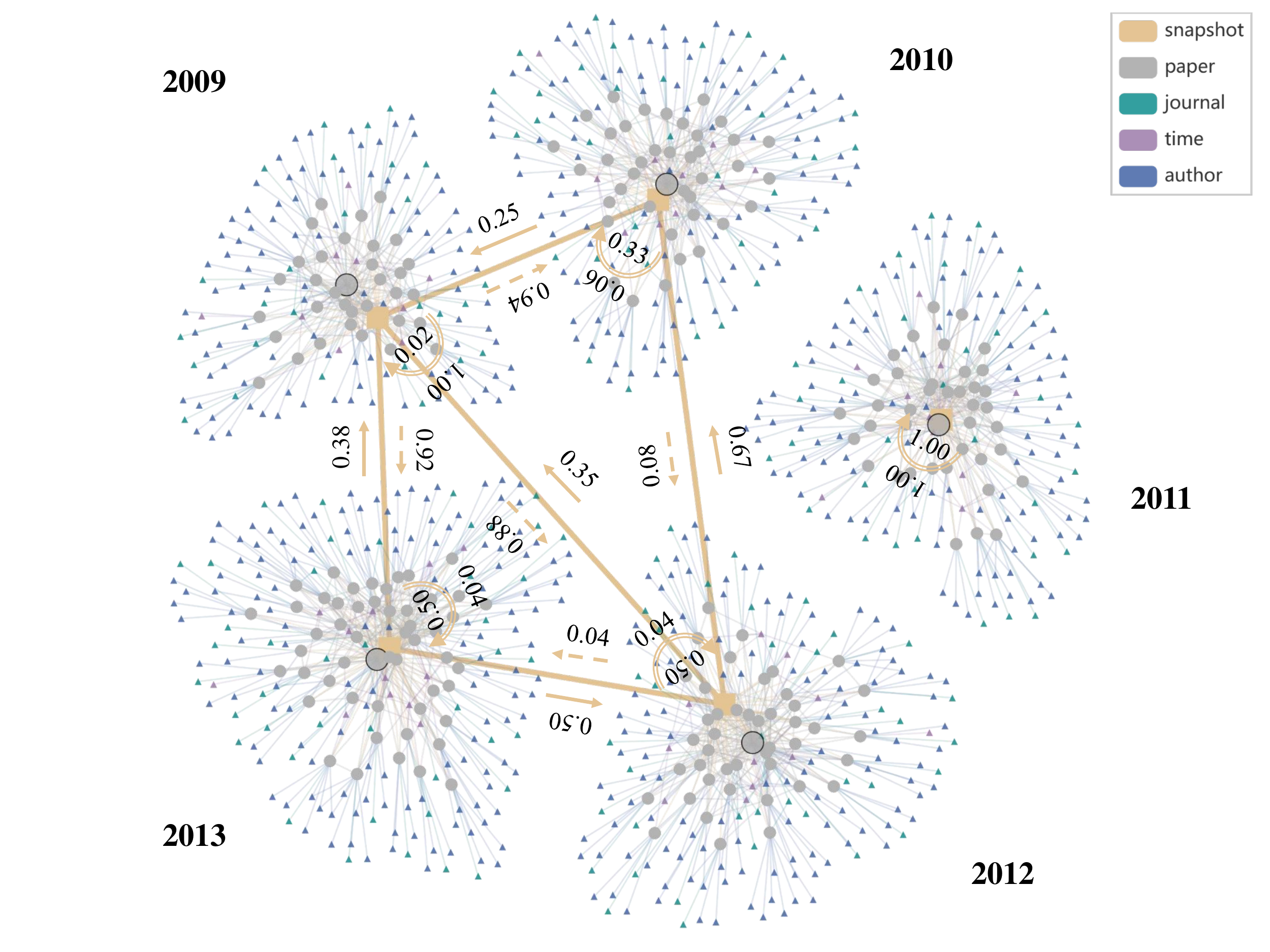}
  \caption{
The Hierarchical and Heterogeneous Graph of the example. The target \textit{paper} nodes are highlighted with black outlines.
  }
\label{fig:whole}
\end{figure*}
To delve into the underlying mechanism of our proposed model, we have selected a challenging sample from DBLP, inaccurately predicted by MUCas, for the detailed case study. This example, published in 2006, is one of the common papers visualized in Section \ref{sec:visual}. As depicted in Figure \ref{fig:whole}, the hierarchical and heterogeneous graph can be divided into five parts, representing different time points. Remarkably, in 2011, this paper received no citations, leading to the absence of links between 2011 and other time points. However, links exist among all other subgraphs and they are influenced to varying extents by different time points, forming edge weights in the temporal graphs. Our model is designed to capture these intricate temporal interactions between different time points through the proposed weighted temporal graphs.

\begin{figure*}[!h]
  \centering
  	\centering
        \subfloat[C-GIN (2013)]{\includegraphics[width=0.432\textwidth]{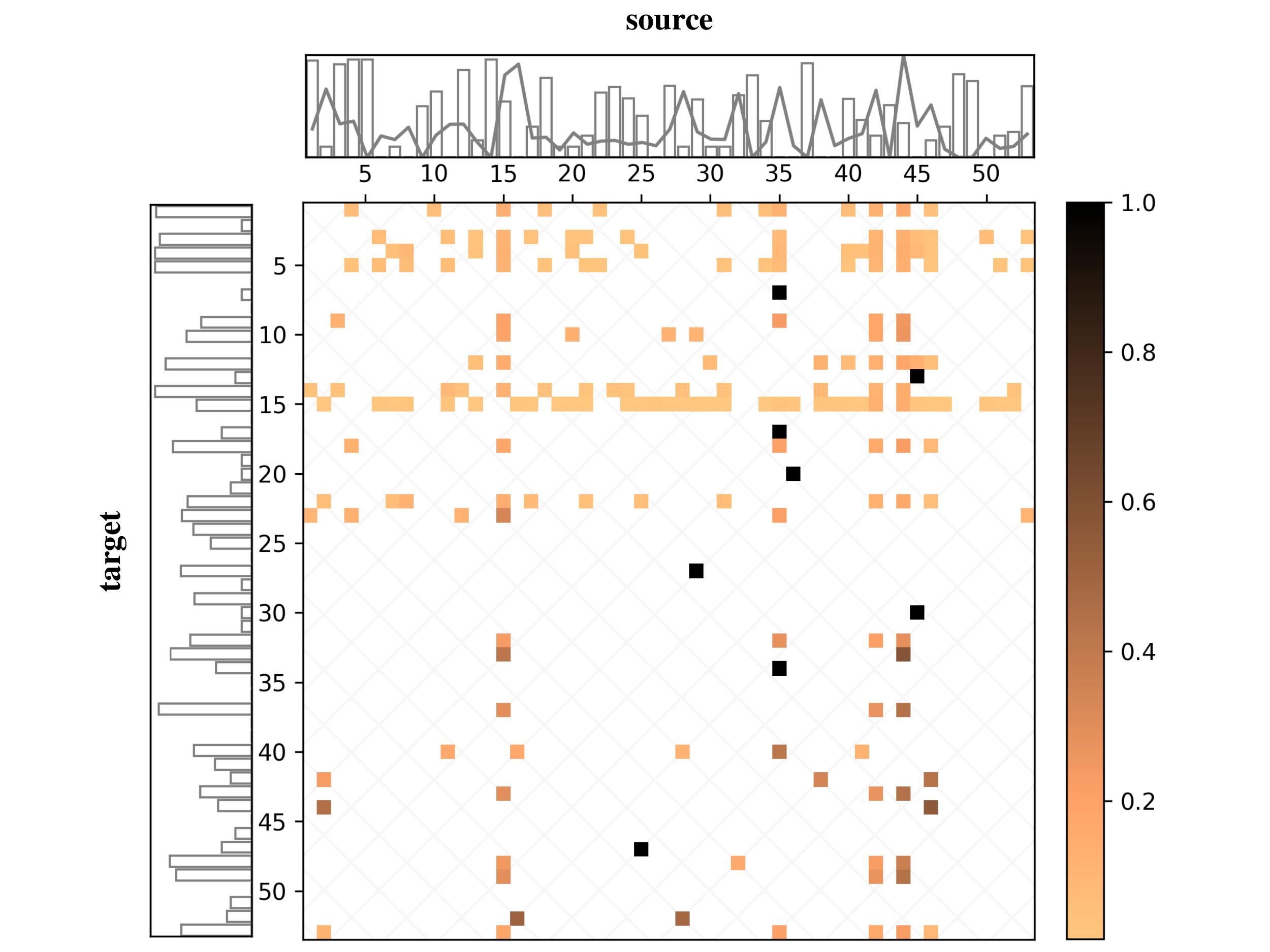}%
        \label{fig:c-gin-d}}
                \hfil
	\subfloat[R-GAT (2009)]{\includegraphics[width=0.432\textwidth]{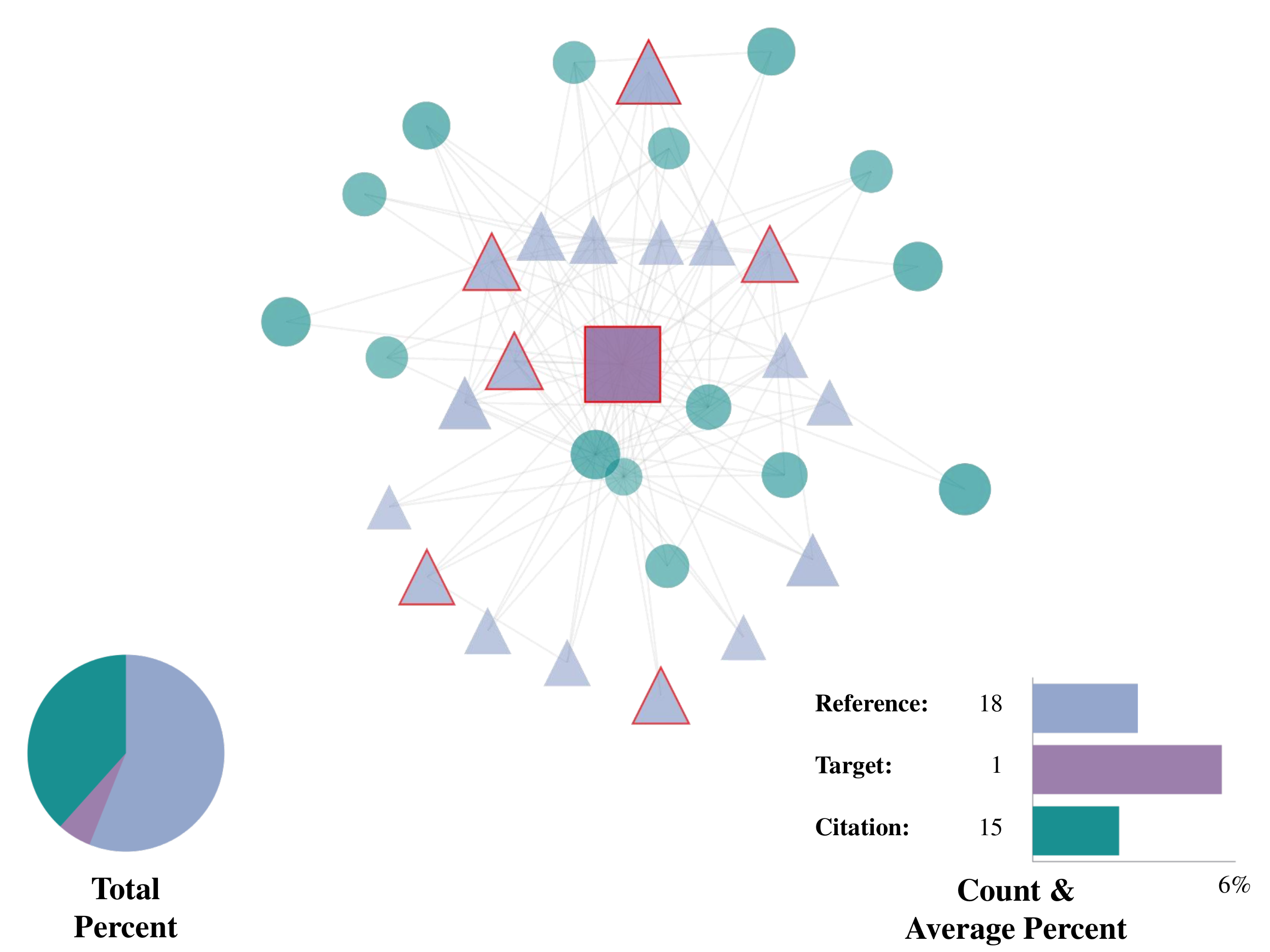}%
		\label{fig:2009}}
        \hfil
	\subfloat[R-GAT (2010)]{\includegraphics[width=0.432\textwidth]{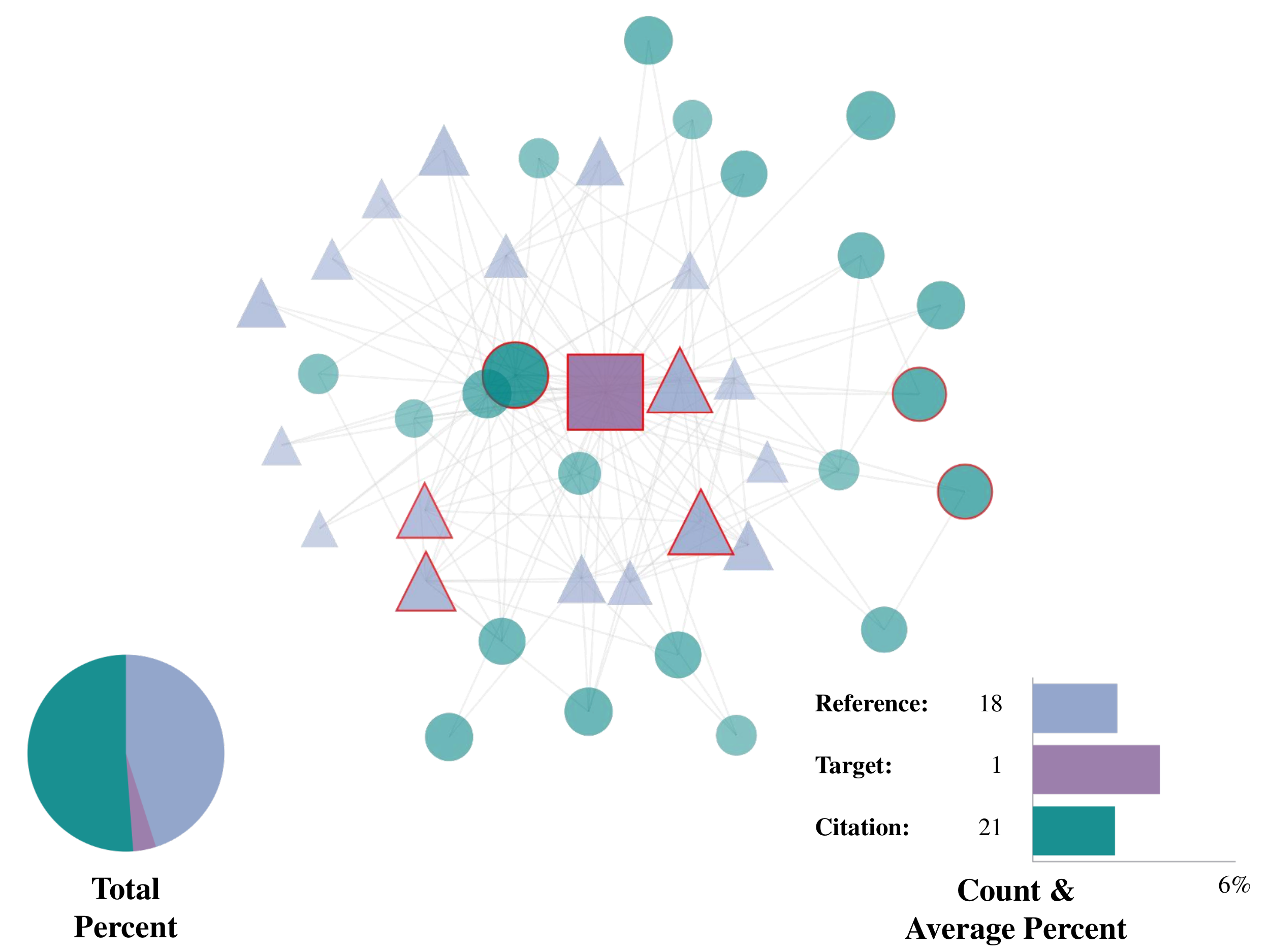}%
		\label{fig:2010}}
        \hfil
	\subfloat[R-GAT (2011)]{\includegraphics[width=0.432\textwidth]{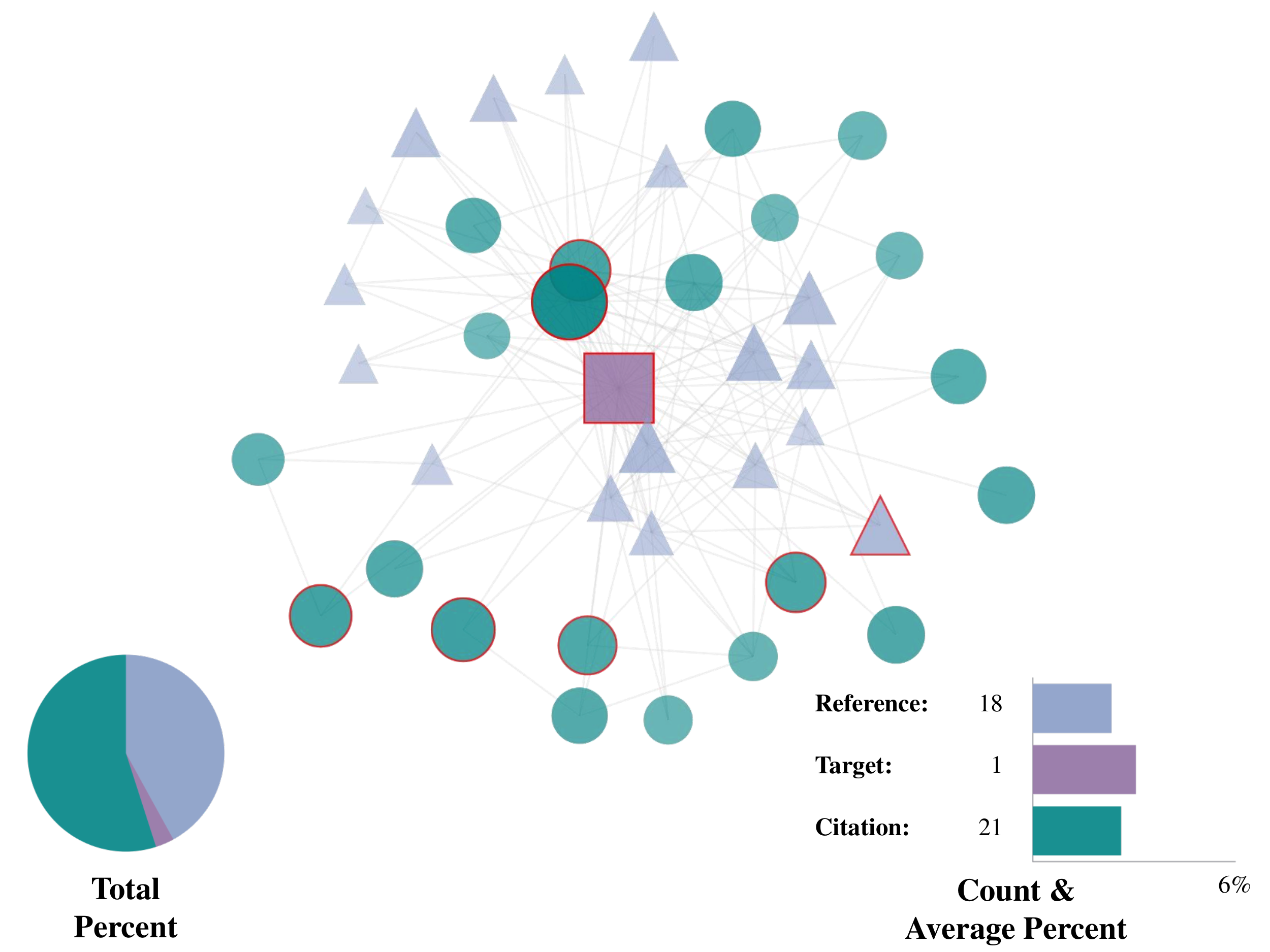}%
		\label{fig:2011}}
                \hfil
	\subfloat[R-GAT (2012)]{\includegraphics[width=0.432\textwidth]{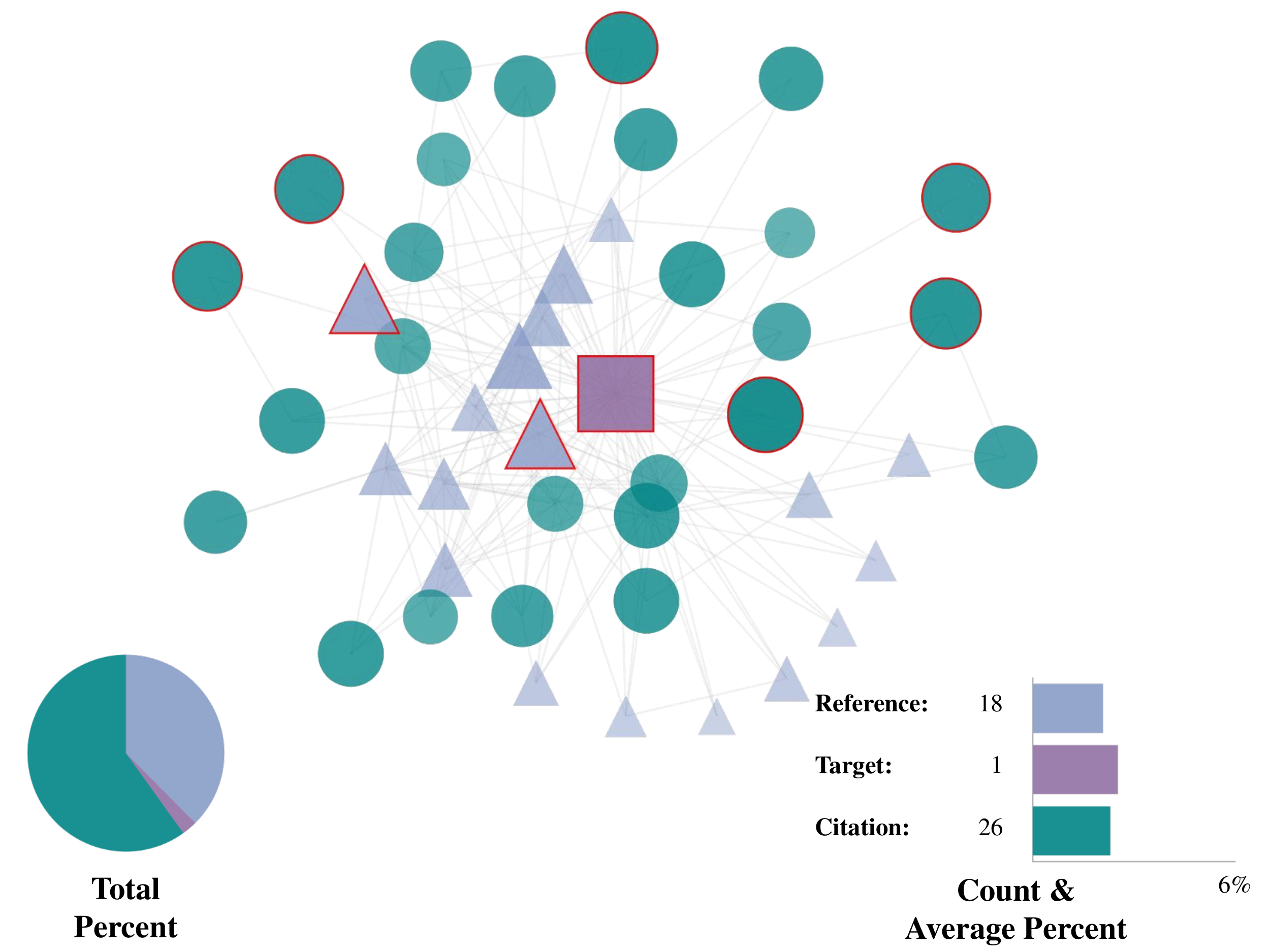}%
		\label{fig:2012}}
          \hfil
	\subfloat[R-GAT (2013)]{\includegraphics[width=0.432\textwidth]{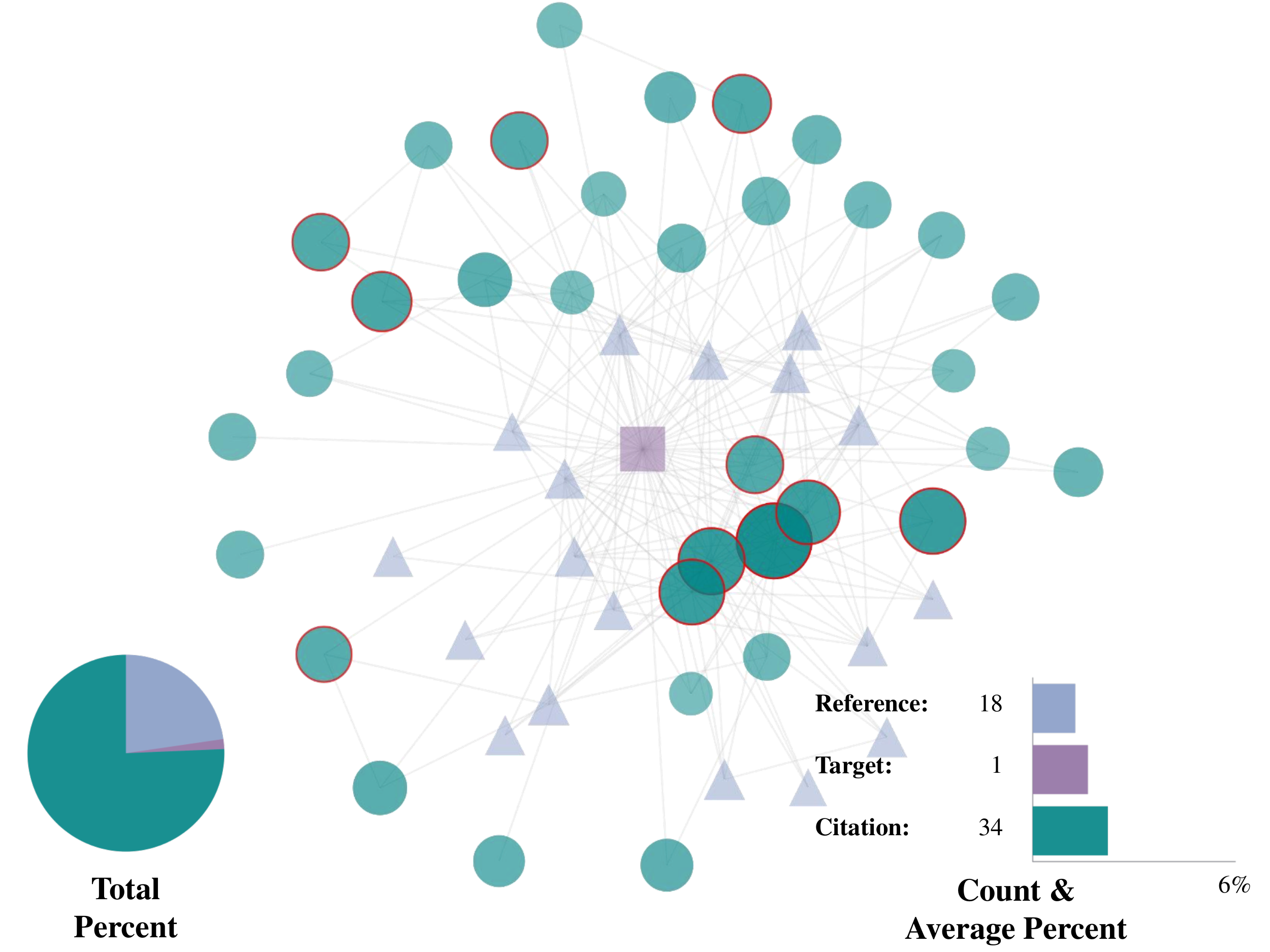}%
		\label{fig:2013}}
    \caption{The node-level detailed weight of proposed C-GIN and R-GAT. (1) For C-GIN, we demonstrate all \textit{paper} node attention weights for the last year (2013) using a heatmap. The aligned small figure shows the globally-received citations, logarithmically scaled with the bar plot, and the weight of source nodes average across all related edges with the line plot. (2) For R-GAT, the graph is split into multiple subgraphs by time, and we focus on the attention weight trend for both \textit{paper} node type and \textit{snapshot} node type. We incorporate opacity and node size to indicate weight differences, with nodes having the top 20\% attention weight highlighted with red outlines. In the bottom-left corner, we show the total percentage of different \textit{paper} node types, while the bottom-right corner presents the count and average percentage of all nodes.}
\label{fig:case_study}
\end{figure*}

For this example, we have visualized the node-level detailed attention weights of the proposed C-GIN and R-GAT models to provide a closer view of their functioning. For C-GIN, the sparsity of ``cites'' in the \textit{paper} subgraphs results in a correspondingly sparse heatmap. As illustrated in Figure \ref{fig:c-gin-d}, despite influential papers having received a large number of citations, they rarely cite other papers as the source nodes. As a result, the citations from remaining highly-cited papers assume crucial significance, underscoring the necessity of C-GIN. Among these papers, the $15_{th}$ paper emerges as the most popular node for both source and target nodes. As the node represents the target paper, concentrating on it is a rational approach for other nodes, thereby showcasing the effectiveness of C-GIN. Furthermore, other highly-cited papers, such as the $32_{th}$ and $44_{th}$, also hold substantial weights. Consequently, C-GIN effectively enables the model to focus on rare, highly-cited papers without dominating the results and maintaining flexibility in attention allocation.

For R-GAT, we partition the entire hierarchical and heterogeneous graph into multiple snapshots to highlight the evolutionary perspective of the proposed model across different time points. During the initial years when the paper is newly published with limited citations, the snapshot gives more attention to the reference and target papers. As shown in Figure \ref{fig:2009}, the proportion of references is above 50\%, and most of the top nodes are references.
Two specific snapshots, 2010 and 2011, require extra focus as they share the same structures but exhibit different attentions. In 2011, the paper has been published for 5 years, thus leading to a change in the type of the \textit{snapshot} node. Therefore, they review the same nodes from different perspectives. In 2010, the newly-published paper still focuses on the reference. Although the number of citations surpasses that of references, the average percentage of references remains higher than that of citations. In contrast, as shown in Figure \ref{fig:2011}, citations exceed references in both the total and average percentages in 2011. The top nodes also shift from references to citations.
Over time, the target paper receives more citations, leading to a larger total percentage of citations. However, the average percentage decreases due to the increased volume. In the final subgraph of Figure \ref{fig:2013}, the average percentage of citations surpasses that of the target paper for the first time, occupying all the top positions.
In conclusion, the detailed node-level attention visualization of the case study aligns with the attention visualization of the total, further confirming the rationale behind R-GAT. It effectively aggregates information from papers within subgraphs determined by both the \textit{paper} node type and the \textit{snapshot} node from an evolutionary perspective.

In summary, the proposed H$^2$CGL enables us to comprehend the dynamics of the citation context by leveraging the constructed Hierarchical and Heterogeneous Graph and the proposed modules. Notably, the C-GIN and R-GAT can effectively encode the distinctive features of the dynamic citation network in an evolutionary manner. This enhancement contributes to the rationality and increased potency of the proposed model.

\section{Conclusion}
In this study, we reformulate the impact prediction task as potential citation count prediction and interval classification. We collect all papers published before the observation point. Then, we design a novel model named Hierarchical and Heterogeneous Contrastive Graph Learning Model (\textbf{H$^2$CGL}) to model citation network dynamics. Particularly, C-GIN could consider highly-cited papers. R-GAT aggregates information to \textit{snapshot} node from references, citations, and the target paper. Weighted GIN enables dynamics modeling as a part of the hierarchical GNN. We further augment the graph learning model via the contrastive learning module. It can make representations of papers that play similar roles in the citation network more sensitive to potential citations. Experimental results on two scholarly datasets demonstrate that H$^2$CGL significantly outperforms alternative approaches for both previous and new papers. Attention visualization demonstrates that H$^2$CGL can reliably make predictions.

In future work, we plan to construct comprehensive datasets encompassing multiple fields of study to simulate more practical scenarios, thereby increasing the practicality and applicability of our approach. More importantly, we aim to design a model that can flexibly handle both papers with accumulated citations and immediately-published ones without any citations, ensuring robustness in real-world applications. Our ultimate objective is to create a model that can be genuinely applied to various applications in reality.

\printcredits

~\\
\noindent
\large
\textbf{Acknowledgement}
\normalsize

This work is supported by the National Natural Science Foundation of China (72204087, 72104212, 71874056, 72234005), the Shanghai Planning Office of Philosophy and Social Science Youth Project (2022ETQ001), the Natural Science Foundation of Zhejiang Province (LY22G030002), the Fundamental Research Funds for the Central Universities. We also appreciate the constructive comments from the anonymous reviewers.

\bibliographystyle{cas-model2-names}

\bibliography{IPM-2023-H2CGL-Ref}

\end{document}